\documentclass[10pt, conference, compsocconf]{IEEEtran}

\usepackage{cite}
\usepackage{amssymb, amsmath, amsthm, amsfonts}
\usepackage[ruled,noend,linesnumbered]{algorithm2e}
\usepackage{algpseudocode}%to use the algorithmic environment to write algorithm pseudocode
\usepackage{graphicx}
\usepackage{subcaption}
\usepackage{wrapfig}
\usepackage{color}
\usepackage{float}
\usepackage{cases}
\usepackage[nodisplayskipstretch]{setspace}
\usepackage{tabularx}
\usepackage{stfloats}
\usepackage{fancyhdr}
\usepackage{listings}
\allowdisplaybreaks
\newtheorem{theorem}{Theorem}
\newtheorem{definition}{Definition}

\newtheorem{proposition}{Proposition}

\newtheorem{property}{Property}
\newtheorem{corollary}{Corollary}

\DeclareMathOperator*{\argmin}{arg\,min}

\usepackage{textcomp}
\def\BibTeX{{\rm B\kern-.05em{\sc i\kern-.025em b}\kern-.08em
    T\kern-.1667em\lower.7ex\hbox{E}\kern-.125emX}}
    
\ifCLASSOPTIONcompsoc

\usepackage[nocompress]{cite}

\else

\usepackage{cite}

\fi

\begin{document}
\title{Differentially Private Model Publishing for Deep Learning}
\author{\IEEEauthorblockN{Lei Yu,~ Ling Liu,~ Calton Pu,~ Mehmet Emre Gursoy, ~Stacey Truex}
\IEEEauthorblockA{School of Computer Science, College of Computing, Georgia Institute of Technology\\
Email: leiyu@gatech.edu, \{ling.liu,calton.pu\}@cc.gatech.edu, \{memregursoy,staceytruex\}@gatech.edu}
}

\pagestyle{empty}
\fancyhf{}
\renewcommand{\headrulewidth}{0pt}
\cfoot{\thepage}

\chead{A preliminary version of this paper appears in the proceedings of 40th IEEE Symposium on Security and Privacy (IEEE S\&P 2019). This version contains a few refinements, corrections and extensions.}
\maketitle
\thispagestyle{fancy}

\begin{abstract}
Deep learning techniques based on neural networks have shown significant success in a wide range of AI tasks.
Large-scale training datasets are one of the critical factors for their success. However, when the training datasets are crowdsourced from individuals and contain sensitive information, the model parameters may encode private information and bear the risks of privacy leakage. The recent growing trend of the sharing and publishing of pre-trained models further aggravates such privacy risks. To tackle this problem, we propose a differentially private approach for training neural networks. Our approach includes several new techniques for optimizing both privacy loss and model accuracy. We employ a generalization of differential privacy called concentrated differential privacy(CDP), with both a formal and refined privacy loss analysis on two different data batching methods. We implement a dynamic privacy budget allocator over the course of training to improve model accuracy. Extensive experiments demonstrate that our approach effectively improves privacy loss accounting, training efficiency and model quality under a given privacy budget.
\end{abstract}


\IEEEpeerreviewmaketitle

\section{Introduction}
In recent years, deep learning techniques based on artificial neural networks have dramatically advanced the state of the art in a wide range of AI tasks such as speech recognition, image classification, natural language processing and game playing. Its success relies on three sources of advancement: high-performance computing, large-scale datasets, and the increasing number of open source deep learning frameworks, such as TensorFlow, Caffe, and Torch.

\noindent
{\bf Privacy Concerns in Deep Learning.}~
However, recent studies on membership attacks and model inversion attacks have exposed potential privacy risks from a number of dimensions. First, large-scale datasets are collected from individuals via crowdsourcing platforms, containing private information such as location, images, medical, and financial data of the users. The users usually do not have any control over how their data is being used or shared once collected. Second, deep neural networks have a large number of hidden layers, leading to a large effective capacity that could be sufficient for encoding the details of some individual's data into model parameters or even memorizing the entire dataset~\cite{ZhangBHRV16}. It has been shown that individual information can be effectively extracted from neural networks~\cite{Fredrikson:2015:MIA,Shokri:MI:2017}. Therefore, there are severe privacy concerns accompanied with the broad deployment of deep learning applications and deep learning as a service .

On the other hand, the publishing and sharing of trained deep learning models has been gaining growing interest. Google's cloud machine learning services provide several pre-trained models usable out-of-the-box through a set of APIs. The model owners can also publish their trained models to the cloud and allow other users to perform predictions through APIs. In mobile applications, entire models are stored on-device to enable power-efficient and low-latency inference. 
Transfer learning~\cite{YosinskiCBL14}, a key technique of deep learning, can leverage and adapt the already existing models to new classes of data, saving the effort of training the entire neural network from scratch.
People who only have small datasets can use the model trained on a large dataset as a fixed feature extractor in their neural networks or adapt the model to their own domain. Transfer learning is believed to be the next driver of machine learning success in industry and will significantly stimulate the sharing of pre-trained models.
A large number of pre-trained models have been publicly available in model zoo repositories~\cite{modelzoo}.
In these cases, the model parameters are entirely exposed, making it easier for adversaries to launch inference attacks, such as membership attacks~\cite{Shokri:MI:2017} or model inversion attacks~\cite{Fredrikson:2015:MIA}, to infer sensitive data records of individuals in the training datasets.
Even by providing only the query APIs to access remote trained models, the model parameters may still be extracted from prediction queries and in turn used to infer the sensitive training data~\cite{stealml}.
Therefore, it is imperative to develop principled privacy-preserving deep learning techniques to protect private training data against adversaries with full knowledge of model parameters.

\noindent
{\bf Deep learning with Differential Privacy.}~
Although privacy-preserving machine learning has attracted much attention over the last decade, privacy preserving deep learning was first proposed in 2015~\cite{Shokri:2015:PDL}. The proposed approach argues for privacy-preserving model training in a collaborative federated learning system and involves multiple participants jointly training a model by sharing sanitized parameters while keeping their training data private and local. The first proposal for deep learning with differential privacy was presented in 2016~\cite{Abadi:2016:DLD}.
Differential privacy (DP), a defacto standard for privacy that offers provable privacy guarantees, has been applied for privacy-preserving machine learning~\cite{Chaudhuri:2008:PLR,Zhang:2012:FMR,Rubinstein2009LearningIA,Chaudhuri:2011:DPE,BassilyST14}. DP characterizes the difference in output between two input datasets differing by at most one element. This characterization is challenging with deep learning because the internal representations of deep neural networks are notoriously difficult to understand. Prior works~\cite{Abadi:2016:DLD, DBLP:journals/corr/abs-1710-06963,ss} suggest using the norm gradient clipping in the stochastic gradient descent (SGD) algorithm to bound the influence of any single example on the gradients and then applying differentially private mechanisms to perturb the gradients accordingly. By ensuring that each gradient descent step is differentially private, the final output model satisfies a certain level of differential privacy given the composition property. It is known that the SGD training process of a deep neural network tends to involve a large number of iterations. Given a target differential privacy guarantee, the differentially private training algorithm needs a tight estimation on the privacy loss for the composition of DP. This is necessary for the algorithm to effectively track cumulative privacy loss during the training process and, if necessary, terminate before the loss exceeds the privacy budget. Unfortunately, Abadi and his co-authors~\cite{Abadi:2016:DLD} have shown the existing strong composition theorem~\cite{advcomposition} for differential privacy does not yield a tight analysis. To address this problem, the moments accountant method is proposed~\cite{Abadi:2016:DLD}, which tracks the log moments of the privacy loss variable and provides a much tighter estimate of the privacy loss for the composition of Gaussian mechanisms under random sampling.

In this paper, however, we examine several issues and possible improvement on using the differentially private SGD (DP-SGD) algorithm and privacy accounting method proposed in~\cite{Abadi:2016:DLD}. The first problem is related to the underestimation of privacy loss caused by data batching methods. For computational efficiency, the SGD algorithm usually takes small batches from the training dataset to iteratively compute gradients and update model parameters. The DP-SGD approach in~\cite{Abadi:2016:DLD} exploits the privacy amplification of random sampling to produce a tighter estimation of privacy loss. This is based on the assumption that the data batches for mini-batch SGD input are generated through random sampling with replacement on the training dataset. In practice, for better efficiency, data batching is implemented via random reshuffling, which randomly shuffles the training dataset and then partitions them into batches of the similar size~\cite{inputpipe,Goodfellow-et-al-2016}. Therefore, it has been implemented as standard APIs in deep learnig frameworks like Tensorflow. Several existing DP-SGD implementations~\cite{Esteban2018RealvaluedT,DBLP:journals/corr/dpdata,tfdpcode, tplib} use moments accountant(MA) method to track privacy loss, while implementing random shuffling or using its APIs to generate data batches which may violate the random sampling assumption required by MA. Table \ref{shuffleexample} demonstrates some examples.
Our anlysis shows that random reshuffling incurs higher privacy loss than random sampling.
Thus, simply treating reshuffling and random sampling as the same will lead to the underestimation of privacy loss and the privacy accounting should depend on how data is accessed.

\begin{table*}
\centering
\caption{We examine a few data batching APIs in Tensorflow. Given a list of numbers from 0 to 9 as input data, we use them to generate batches of size 1 and show the output batch for each training iteration in an epoch. In addition, we run the true random sampling that samples each number with probability batch\_size/data\_size=0.1 to generate batches in 10 iterations. The source code of our test is given in Appendix \ref{ap:batch}.
}
\small
\label{shuffleexample}
\begin{tabular}{l|l|l} 
\hline
Batching methods      & Input data              & example of output batches in one epoch     \\ 
\hline
tf.train.shuffle\_batch    & [0,1,2,\ldots,9] & [2],[6],[8],[0],[5],[7],[1],[4],[9],[3]  \\ 
\hline
tf.estimator.inputs.numpy\_input\_fn  & [0,1,2,\ldots,9] & [5],[1],[4],[9],[8],[3],[7],[2],[6],[0]  \\
\hline
random sampling with prob=0.1  & [0,1,2,\ldots,9] & [2], [1 7], [] ,[4 5], [], [7] ,[5 7 9] ,[], [1], [1 3 7]  \\
\hline
\end{tabular}
\end{table*}

The second issue is the need for a tight analysis on cumulative privacy loss for DP-SGD which tends to have a large number of iterations. To address this problem, we propose the use of concentrated differential privacy (CDP), a generalization of differential privacy recently introduced by Dwork and Rothblum~\cite{Dwork2016ConcentratedPrivacy}. CDP focuses on cumulative privacy loss for a large number of computations and provides a sharper analysis tool. Based on CDP, we analyze the privacy loss under random reshuffling and random sampling respectively. In particular, our CDP based privacy accounting method for random reshuffling provides a tighter estimation of privacy loss  even than the result produced by the strong composition theorem with having privacy amplification effect of random sampling.
In addition, we also consider using CDP for the privacy loss analysis under random sampling based data batching. We find that CDP is not able to capture the privacy amplification effect of random sampling. Accordingly, we propose to consider a relaxation of CDP and convert it to traditional $(\epsilon,\delta)$-differential privacy. Our method provides a slightly loose but easy method to approximately estimate the privacy loss for random sampling.

The third novelty of our approach to differentially private deep learning is our development of a dynamic privacy budget allocation to improve model accuracy under differentially private training. The perturbed gradients during the training process inevitably degrade model accuracy. We aim to provide differential privacy guarantees on the final output models. This means that we are much less concerned with the privacy loss of a single iteration. This provides opportunities to optimize the model accuracy via adjusting privacy budget allocation for every training iteration. In this paper, we propose a set of dynamic privacy budget allocation methods and our extensive experiments demonstrate benefits for improving the model accuracy. It is worth noting that the techniques proposed in this paper not only apply to neural networks but also may apply to any other iterative learning algorithms.

The remainder of the paper is as follows. We review necessary background in Section \ref{sec:background}, provide an overview of our approach in Section \ref{sec:overview}, and then give a detailed technical development in Section \ref{sec:details}. Section \ref{sec:exp} describes experimental results. Some discussions are given in Section \ref{sec:discussion}. Section \ref{sec:related} presents related work and Section \ref{sec:conclude} concludes the paper. Deferred proofs are provided in the Appendix.

\section{BACKGROUND}
\label{sec:background}
\subsection{Deep Learning}
Deep learning uses neural networks that are defined as a hierarchical composition of parameterized functions to model the input data. For supervised learning, the training data are labeled with correct classes, and a multi-layer neural network is deployed to model the correlation between data instances and their labels. 
A typical neural network consists of $n(n>1)$ layers of neurons. Each layer of neurons is parameterized by a weight matrix $W^{(l)}$ and a bias vector $b^{(l)}$. Layers apply an affine transformation to the previous layer's output and then computes an activation function $\sigma$ over that. Typical examples of the activation function $\sigma$ are sigmoid, rectified linear unit(ReLU) and tanh.

The training of a neural network aims to learn the parameters $\theta=\{W^{(l)},b^{(l)}|1 \le l \le n\}$ that minimize a loss function $L$ defined to represent the penalty for misclassifying the training data. It is usually a non-convex optimization problem and solved by gradient descent. The gradient descent method iteratively computes the gradient of the loss function $L$ and updates the parameters every step until the loss converges to a local optimum. In practice, the training of neural networks uses the mini-batch stochastic gradient descent (SGD) algorithm, which is much more efficient for large datasets. At each step a batch $B$ of examples is sampled from the training dataset and the gradient of the average loss is computed, i.e., $\frac{1}{|B|} \sum_{x\in B} \nabla _\theta L(\theta, x)$ as $\nabla _\theta L(\theta)$. The SGD algorithm then applies the following update rule for parameters $\theta$ 
\begin{equation}
\label{eq:gradient}
\theta = \theta - \alpha \nabla _\theta L(\theta)
\end{equation}
where $\alpha$ is the learning rate.
The running time of the mini-batch SGD algorithm is usually expressed as the number of \emph{epochs}. Each epoch consists of all of the batches of the training dataset, i.e., in an epoch every example has been seen once.
Within an epoch, the pass of one batch of examples for updating the model parameters is called one \emph{iteration}.

\subsection{Differential Privacy}
Differential privacy is a rigorous mathematical framework that formally defines the privacy properties of data analysis algorithms. Informally it requires that any changes to a single data point in the training dataset can only cause statistically insignificant changes to the algorithm's output.
\begin{definition}[Differential Privacy~\cite{dip}]
A randomized mechanism $\mathcal{A}$ provides $(\epsilon, \delta)$-differential privacy if for any two neighboring database $D$ and $D'$ that differ in only a single entry, $\forall S \subseteq Range(\mathcal{A})$,
\begin{equation}
\Pr(\mathcal{A}(D) \in S) \le e^\epsilon \Pr(\mathcal{A})(D') \in S) + \delta
\end{equation}
\end{definition}
If $\delta=0$, $\mathcal{A}$ is said to be $\epsilon$-differential privacy. In the rest of this paper, we write $(\epsilon, \delta)$-DP for short.

The standard approach to achieving differential privacy is the sensitivity method~\cite{Dwork:2006:CNS,dip} that adds to the output some noise that is proportional to the sensitivity of the query function. The sensitivity measures the maximum change of the output due to the change of a single database entry.
\begin{definition}[Sensitivity~\cite{Dwork:2006:CNS}]
The sensitivity of a query function $q: \mathcal{D} \rightarrow \mathbb{R}^d$ is
\begin{equation}
\Delta = \max\limits_{D,D'} ||q(D)-q(D') ||
\end{equation}
where $D$, $D' \in \mathcal{D}$ are any two neighboring datasets that differ at most one element, $||\cdot||$ denotes $L_1$ or $L_2$ norm.
\end{definition}
In this paper, we choose the Gaussian mechanism that uses $L_2$ norm sensitivity. It adds zero-mean Gaussian noise with variance $\Delta ^2 \sigma^2$ in each coordinate of the output $q(D)$, as
\begin{equation}
q(D)+\mathcal{N}(0, \Delta ^2 \sigma^2 \mathbf{I})    
\end{equation}
It satisfies ($\epsilon$,$\delta$)-DP if $\sigma^2>$ $2\log(\frac{1.25}{\delta})/\epsilon^2$ and $\epsilon\in$ $(0,1)$~\cite{Dwork:2014:AFD}.

\subsection{Concentrated Differential Privacy}
Concentrated differential privacy (CDP) is a generalization of differential privacy recently introduced by Dwork and Rothblum~\cite{Dwork2016ConcentratedPrivacy}. It aims to make privacy-preserving algorithms more practical for large numbers of computations than traditional DP while still providing strong privacy guarantees. It allows the computation to have much less concern about single-query loss but high probability bounds for the cumulative loss, and provides sharper and more accurate analysis on the cumulative loss for multiple computations compared to the popular ($\epsilon$, $\delta$)-DP. %Because the training in deep learning involves a large number of iterations, CDP is an appropriate choice for designing differentially private training algorithms. 

CDP considers privacy loss on an outcome $o$ as a random variable when the randomized mechanism $\mathcal{A}$ operates on two adjacent database $D$ and $D'$:
\begin{equation}
\label{eq:privacyloss}
L^{(o)}_{(\mathcal{A}(D)||\mathcal{A}(D'))}  \overset{\Delta}{=} \log \frac{\Pr(\mathcal{A}(D)=o)}{\Pr(\mathcal{A}(D')=o)}
\end{equation}
The $(\epsilon, \delta)$-DP guarantee ensures that the privacy loss variable is bounded by $\epsilon$ but exceeds that with probability no more than $\delta$. As a relaxation to that, $(\mu, \tau)$-concentrated differential privacy, $(\mu, \tau)$-CDP for short~\cite{Dwork2016ConcentratedPrivacy}, ensures that the mean (i.e., expectation) of the privacy loss is no more than $\mu$ and the probability of the loss exceeding its mean by an amount of $t \cdot \tau$ is bounded by $e^{-t^2/2}$. An alternative formulation of CDP to Dwork and Rothblum's $(\mu, \tau)$-CDP is proposed by Bun and Steinke~\cite{BunS16}, called ``zero-concentrated differential privacy'' (zCDP for short). Instead of mean concentrated as $(\mu, \tau)$-CDP, zCDP makes privacy loss concentrated around zero (hence the name), still following sub-Gaussian such that larger deviations from zero become increasingly unlikely.
\begin{definition}[Zero-Concentrated Differential Privacy (zCDP)\cite{BunS16}]
A randomized mechanism $\mathcal{A}$ is $\rho$-zero concentrated differentially private (i.e., $\rho$-zCDP) if for any two neighboring databases $D$ and $D'$ that differ in only a single entry and all $\alpha \in (1, \infty)$,
\begin{equation}
\label{eq:zcdp}
D_\alpha (\mathcal{A}(D)||\mathcal{A}(D')) \overset{\Delta}{=} \frac{1}{\alpha -1} \log \big(\mathbb{E}\left[ e^{(\alpha-1)L^{(o)}} \right] \big) \le \rho \alpha 
\end{equation}
Where $D_\alpha (\mathcal{A}(D)||\mathcal{A}(D'))$ called $\alpha$-R\'enyi divergence between the distributions of $\mathcal{A}(D)$ and $\mathcal{A}(D')$.
\end{definition}

The $(\epsilon, \delta)$-DP bounds the privacy loss by ensuring $\Pr(L^{(o)}> \epsilon) \le \delta$. In contrast, zCDP entails a bound on the moment generating function of privacy loss $L^{(o)}$, indicated by an equivalent form of (\ref{eq:zcdp})
\begin{equation}
\label{eq:zcdpeq}
\mathbb{E}\left[e^{(\alpha-1) L^{(o)}} \right] \le e^{(\alpha-1)\alpha \rho}
\end{equation}
This implies that for zCDP, privacy loss $L^{(o)}$ is assumed to be a sub-Gaussian random variable such that it has a strong tail decay property, namely, $\Pr(L^{(o)} > t+\rho) \le e^{-t^2/(4\rho)}$ for all $t>0$ ~\cite{BunS16}. The following propositions are restatements of some zCDP results given in \cite{BunS16} that will be used in our paper.

\begin{proposition}
\label{prop:zcdpvsdp}
If $\mathcal{A}$ provides $\rho$-zCDP, then $\mathcal{A}$ provides $(\rho+2\sqrt{\rho \log(1/\delta)}, \delta)$-DP for any $\delta>0$.
\end{proposition}
\begin{proposition}
\label{prop2:Gaussianzcdp}
The Gaussian mechanism with noise $\mathcal{N}(0, \Delta ^2 \sigma^2 \mathbf{I})$ satisfies ($\frac{1 }{2\sigma^2}$)-zCDP.
\end{proposition}

We use zCDP instead of original $(\mu, \tau)$-CDP because zCDP is comparable to $(\epsilon, \delta)$-DP, as indicated by Proposition \ref{prop:zcdpvsdp} and is immune to post-processing while $(\mu, \tau)$-CDP is not closed under post-processing~\cite{BunS16}. 

\subsection{Composition}
Differential privacy offers elegant composition properties that enable more complex algorithms and data analysis task via the composition of multiple differentially private building blocks. The composition should have privacy guarantees degraded gracefully with multiple outputs that may be subjected to the joint analysis from building blocks.

For a sequential composition of $k$ mechanisms $\mathcal{A}_1, \ldots, \mathcal{A}_k$ satisfying $(\epsilon_i, \delta_i)$-DP for $i$=1,$\ldots$, $k$ respectively, the basic composition result~\cite{Dwork:2014:AFD} shows that the privacy composes linearly, i.e., the sequential composition satisfies $(\sum_i^k\epsilon_i, \sum_i^k\delta_i)$-DP. When $\epsilon_i=\epsilon$ and $\delta_i=\delta$, the strong composition bound from~\cite{advcomposition} states that the composition satisfies ($\epsilon \sqrt{2klog(1/\delta')}+k\epsilon(e^\epsilon-1)$, $k\delta+\delta'$)-DP. For zCDP, it has a simple linear composition property~\cite{BunS16}:
\begin{theorem}
\label{prop:Gaussianzcdp}
Two randomized mechanisms $\mathcal{A}_1$ and $\mathcal{A}_2$ satisfy $\rho_1$-zCDP and $\rho_2$-zCDP respectively, their sequential composition  $\mathcal{A} = (\mathcal{A}_1,\mathcal{A}_2)$ satisfies ($\rho_1+\rho_2$)-zCDP.
\end{theorem}

Compared with $(\epsilon, \delta)$-DP, CDP provides a tighter bound on the cumulative privacy loss under composition, which makes it more suitable for algorithms running a large number of iterations. In other words, while providing the same privacy guarantee, CDP allows lower noise scale and thus better accuracy. Consider $k$ iterative composition of a Gaussian mechanism with noise $\mathcal{N}(0, \sigma^2 \mathbf{I})$. To guarantee the final ($\epsilon$, $\delta$)-DP, in terms of $(\epsilon_i, \delta_i)$-DP for every iteration, the permitted loss $\epsilon_i$ of each iteration is $\epsilon_i = \epsilon/(2\sqrt{2k \log(1/(\delta- k\delta_i))})$, which will be very low when $k$ is large. The noise scale is $\sigma = \frac{\sqrt{k}}{\epsilon}(4\sqrt{\log(1.25/\delta_i)\log(1/(\delta- k\delta_i))})$. Suppose $\delta_i = \delta / (k+1)$, we then have $\sigma > \frac{\sqrt{k}}{\epsilon}(4\log((k+1)/\delta))$.
In contrast, with using $\rho_i$-zCDP for every iteration, because $\rho \approx \epsilon^2 /(4\log(1/\delta))$ zCDP satisfies ($\epsilon$, $\delta$)-DP~\cite{BunS16} and $\rho_i= \frac{1}{k} \rho$. It is easy to show that the noise scale $\sigma = \frac{\sqrt{k}}{\epsilon}\sqrt{2log (1/\delta)}$ which is multiple times smaller than the noise scale derived under $(\epsilon, \delta)$-DP. On the other hand, a single parameter $\rho$ of zCDP and its linear composition naturally fit the concept of a privacy budget. Thus, zCDP is an appropriate choice for privacy accounting.

\section{Overview}
\label{sec:overview}
Because it is difficult to characterize the maximum difference of the model parameters over any two neighboring datasets for neural networks, differentially private deep learning~\cite{Abadi:2016:DLD, DBLP:journals/corr/abs-1710-06963,ss} relies on differentially private stochastic gradient descent (DP-SGD) to control the influence of training data on the model. This approach explicitly bounds per-example gradients $\nabla _\theta L(\theta, x)$ in every iteration by clipping the $L_2$ norm of gradient vectors. Given a clipping threshold $C$, this is done by replacing the gradient vector $\mathbf{g}$ with $\mathbf{g}/\max(1, \frac{||\mathbf{g}||_2}{C})$ which scales $\mathbf{g}$ down to norm $C$ if $||\mathbf{g}||_2>C$.
A Gaussian mechanism with $L_2$ norm sensitivity of $C$ is then applied to perturb the gradients before the gradient descent step in Eq.~(\ref{eq:gradient}) updates the model parameters.
Because each SGD step is differentially private, by the composition property of differential privacy, the final model parameters are also differentially private. The problem with DP-SGD is that the training of a deep neural network (DNN) tends to have a large number of iterations, which causes large cumulative privacy loss at the end. Therefore, a tight estimation of privacy loss under composition is critical for allowing lower noise scale or more training iterations (for desired accuracy) when we have a fixed privacy budget.

To analyze the cumulative privacy loss of DP-SGD, we employ concentrated differential privacy(CDP) which was developed to accommodate a larger number of computations and provides sharper and tighter analysis of privacy loss than the strong composition theorem of $(\epsilon, \delta)$-DP. One way to track the privacy loss of DP-SGD is the Moments Accountant (MA) method proposed by Abadi et al.~\cite{Abadi:2016:DLD}. It assumes that the data batches for mini-batch SGD are generated by randomly sampling examples from the training dataset with replacement, MA takes advantage of the privacy amplification effect of random sampling to achieve a much tighter estimate on privacy loss than the strong composition theorem. It has been shown in~\cite{Li:2012:SAD} that running an $(\epsilon, \delta)$ differentially private mechanism over a set of examples each of which is independently sampled with probability $q$ ($0<q<1$) achieves $(\log(1+q(e^\epsilon-1)), q\delta)$-DP.
However, in practice, random batches are generated by randomly shuffling examples and then partitioning them into batches for computation efficiency, which is distinct from random sampling with replacement. By analyzing the privacy loss under these two data batching methods, random sampling with replacement and random reshuffling respectively, we show that 1) random sampling with replacement and random reshuffling result in different privacy loss; and 2) privacy accounting using the MA method underestimates the actual privacy loss of their neural network training, because it simply regards random reshuffling as random sampling with replacement. To address these problems, we develop a privacy accounting method for random reshuffling, and our algorithm makes proper choices depending on which method is used for data batching. For privacy accounting under random sampling based batching, we show that CDP is unable to capture the privacy amplification effect of random sampling. To address that, we propose a relaxation of zCDP and convert it to $(\epsilon, \delta)$-DP. Compared with MA, our approach provide an explicit expression to approximately estimate the privacy loss. It is easy to compute and can be useful when the users need to decide proper training time, noise scale and sampling ratio during the planning phase.

In our approach, dynamic privacy budget allocation is applied to DP-SGD to improve the model accuracy. In model publishing the privacy loss of each learning step is not our primary concern. This allows us to allocate different privacy budgets to different training epochs as long as we maintain the same overall privacy guarantee. Our dynamic budget allocation approach is in contrast to the previous work~\cite{Abadi:2016:DLD}, which employs a uniform privacy budget allocation, and uses the same noise scale in each step of the whole training process. Our dynamic privacy budget allocation approach leverages several different ways to adjust the noise scale. Our experimental results demonstrate that this approach achieves better model accuracy while retaining the same privacy guarantee.

Algorithm \ref{alg:dpsgd} presents our DP-SGD algorithm. In each iteration, a batch of examples is sampled from the training dataset and the algorithm computes the gradient of the loss on the examples in the batch and uses the average in the gradient descent step. The gradient clipping bounds per-example gradients by $l_2$ norm clipping with a threshold $C$. The Gaussian mechanism adds random noise $\mathcal{N}(0,\sigma_t^2 C^2\mathbb{I})$ to $\sum_{i}\hat{\mathbf{g}}_t(x_i)$ to perturb the gradients in every iteration.
We have a total privacy budget $\rho_{total}$ and cumulative privacy cost $c^{priv}_t$. The way to update $c^{priv}_t$ depends on which batching method is used. If the privacy cost exceeds the total budget, then the training is terminated. In the pseudo code, the function $AdpBudgetAlloc(\rho_{total}, t, T, schedule)$ is used to obtain the noise scale $\sigma_t$ for the current training step $t$, according to $schedule$ that decides how the noise scale is adjusted during the training time. 

\begin{algorithm}[t]
\small
 \caption{Differentially Private SGD Algorithm}
  \label{alg:dpsgd}
  \KwIn{Training examples $\{x_1,\ldots,x_N\}$, learning rate $\eta_t$, group size $L$, gradient norm bound $C$, total privacy budget $\rho_{total}$}
  Initialize $w_0$ \;
  Initialize cumulative privacy loss $c^{priv}_t=0$\;
  \For {$t=1~:~T$}
  {
    \textbf{Dynamic privacy budget allocation}:\\
    $\sigma_t \leftarrow $ $AdpBudgetAlloc(\rho_{total}, t, T, schedule)$\;
    update $c^{priv}_t$ according to data batching method, $t$ and $\sigma_t$\;
    If $c^{priv}_t > \rho_{total}$, break \;
    \textbf{data batching}:\\
    Take a batch of data samples $\mathbb{B}_t$ from the training dataset\;
    $B=|\mathbb{B}_t|$\;
    \textbf{Compute gradient}:\\
    For each $i\in \mathbb{B}_t$, $\mathbf{g}_t(x_i) \leftarrow \bigtriangledown_{w_t} \mathcal{L}(w_t,x_i)$\;
    \textbf{Clip gradient}:\\
    $\hat{\mathbf{g}}_t(x_i) \leftarrow \mathbf{g}_t(x_i)/max\left(1,\frac{||\mathbf{g}_t(x_i)||_2}{C}\right)$\;
    \textbf{Add noise}:\\
    $\widetilde{\mathbf{g}}_t \leftarrow \frac{1}{B} \big(\sum_{i}\hat{\mathbf{g}}_t(x_i)+\mathcal{N}(0,\sigma_t^2C^2\mathbb{I})\big)$\;
    \textbf{Descent}:\\
    $w_{t+1} \leftarrow w_{t}- \eta_t \widetilde{\mathbf{g}}_t$ \;
    
  }
  Output $w_T$ \;
\end{algorithm}

\section{Details of Our Approach}
\label{sec:details}
In this section, we present the details of our approach for differentially private deep learning. We first propose our dynamic privacy budget allocation techniques and then develop privacy accounting methods based on zCDP for different data batching methods.

\subsection{Dynamic Privacy Budget Allocation}
\label{ssec:optbudget}
In Algorithm \ref{alg:dpsgd}, the privacy budget allocated to an epoch decides the noise scale of Gaussian mechanism used by each iteration within that epoch. For a given privacy budget $\rho_{total}$, the final model accuracy depends on how the privacy budget is distributed over the training epochs. Our approach aims to optimize the budget allocation over the training process to obtain a differentially private DNN model with better accuracy.

Concretely, our dynamic privacy budget allocation follows the idea that, as the model accuracy converges, it is expected to have less noise on the gradients, which allows the learning process to get closer to the local optimal spot and achieve better accuracy. A similar strategy has been applied to the learning rate of DNNs in common practice. It is often recommended to reduce the learning rate as the training progresses, instead of using a constant learning rate throughout all epochs, to achieve better accuracy~\cite{decaylearningrate,Chollet2017XceptionDL}.  
%The reason is that, starting with a higher learning rate, SGD can reduce the loss faster, and later decaying the learning rate over time helps SGD to avoid bouncing around, tune and finally settle down into the local optimal spot. Given this idea, we argue that the effects of noise at different training stages on the learning process are different. 
Therefore, we propose a set of methods for privacy budget allocation, which effectively improve the model accuracy by dynamically reducing the noise scale over the training time (as demonstrated by our experiments).

\subsubsection{Adaptive schedule based on public validation dataset}
One approach for adjusting the noise scale is to monitor the validation error during training and reduce the noise scale whenever the validation error stops improving. We propose an adaptive privacy budget allocation that dynamically reduces the noise scale according to the validation accuracy. Every time when the validation accuracy improves by less than a threshold $\delta$, the noise scale is reduced by a factor of $k$ until the total privacy budget runs out. However, when the validation dataset is sampled from the private training dataset, the schedule has some dependency on the private dataset, which adds to the privacy cost. In this case, if we can leverage a small publicly available dataset from the same distribution and use it as our validation dataset, then it will not incur any additional privacy loss.

In our approach, with a public validation dataset, the validation accuracy is checked periodically during the training process to determine if the noise scale needs to be reduced for subsequent epochs. The epochs over which the validation is performed are referred to as validation epochs.
Let $\sigma_{e}$ be the noise scale for the DP-SGD training in the validation epoch $e$, and $S_e$ be the corresponding validation accuracy .
The noise scale for the subsequent epochs is adjusted based on the accuracy difference between current epoch $e$ and the previous validation epoch $e-1$. Initially, $S_0=0$,
\begin{numcases}{\sigma'_e =}
k \sigma_{e}, \text{~~if~}  S_e-S_{e-1} \le \delta \label{eq:vld} \\
\sigma_{e}
\end{numcases}
The updated noise scale $\sigma'_e$ is then applied to the training until the next validation epoch $e+1$.
The above equations amount to say that if the improvement on the validation accuracy $S_e-S_{e-1}$ is less than the threshold $\delta$, it triggers the decay of noise scale $\sigma_e= k \sigma_{e}$ where $k$ ($0<k<1$) is decay rate, a hyperparameter for the schedule. We note that the validation accuracy may not increase monotonically as the training progresses, and its fluctuations may cause unnecessary reduction of noise scale and thus waste on the privacy budget. This motivates us to use the moving average of validation accuracy to improve the effectiveness of validation-based noise scale adjustment: at validation epoch $e$, we define an averaged validation accuracy $\bar{S}_e$ over the previous $m$ validation epochs from $e$, including itself, as follows: 
\begin{equation}
\small
\bar{S}_e = \frac{1}{m}\sum_{i=e-m+1}^{e}S_{i}
\end{equation}
The schedule checks the averaged validation accuracy every $period$ ($period\ge m$) number of validation epochs and compares the current result with that of the last checking time to decide if the noise scale needs to be reduced according to Eq. (\ref{eq:vld}).

\subsubsection{Pre-defined schedules}
When the public validation dataset is not available, we propose to use an alternative approach that pre-defines how the noise scale decreases over time without accessing any datasets or checking the model accuracy.
Concretely, in our approach, the noise scale is reduced over time according to some decay functions. The decay functions update the noise scale by epoch, while the noise scale keeps the same for every iteration within an epoch. By leveraging the Gaussian mechanism that adds noise from $\mathcal{N}(0,\sigma_t^2C^2\mathbb{I})$, we provide four instances of decay functions, all of which take the epoch number $t$ as an argument.

\noindent\emph{a) Time-Based Decay}: It is defined with the mathematical form $$\sigma_t=\sigma_0/(1+kt)$$ where $\sigma_0$ is the initial noise parameter for $\sigma$, $t$ is the epoch number and $k$ ($k>0$) is decay rate.  
when $k<1$, it is known as ``search-then-converge''~\cite{Darken:1990:NLR:118850}, which decreases the noise scale linearly during the SGD search phase when $t$ is less than the ``search time'' $1/k$, and decreases the noise by $1/t$ when $t$ is greater than $1/k$.

\noindent\emph{b) Exponential Decay}: It has the mathematical form $$\sigma_t=\sigma_0 e^{-kt}$$ where $k$ $(k>0)$ is decay rate.

\noindent\emph{c) Step Decay}: Step decay reduces the learning rate by some factor every few epochs. The mathematical form is $$\sigma_t=\sigma_0*k^{ \lfloor t/period \rfloor}$$ where $k$ ($1>k>0$) is the decay factor and $period$ decides how often to reduce noise in terms of the number of epochs. 

\noindent\emph{d) Polynomial Decay}: It has the mathematical form $$\sigma_t=(\sigma_0 - \sigma_{end})*(1-t/period)^k + \sigma_{end}$$ where $k(k>0)$ is the decay power and $t<period$. A polynomial decay function is applied to the initial $\sigma_0$ within the given number of epochs defined by $period$ to reach $\sigma_{end}$ ($\sigma_{end}<\sigma_0$). When $k=1$, this is a linear decay function.

The schedules for noise decay are not limited to the above four instances. In this paper we choose to use these four because they are simple, representative, and also used by learning rate decay in the performance tuning of DNNs. 
Users can apply them in various ways. For example, a user can use them in the middle of training phase, but keep constant noise scale before and after. Note that in the time-based decay and exponential decay, $t$ can be replaced by $\lfloor t/period \rfloor$ as done in the step decay such that the decaying is applied every $period$ number of epochs. The polynomial decay requires a specific end noise scale after $period$ epochs, and we make the noise scale constant after the training time exceed $period$ epochs. 

%It is also worth to note that budget allocation schedules are independent of the learning rate. The budget allocation schedules can work with different learning rate schedules or adaptive learning rate methods. Considering $w_{t+1} \leftarrow w_{t}- \eta_t \widetilde{g}_t$, the budge allocation schedules decide the noise scale for perturbing $\widetilde{g}_t$, and the learning rate $\eta_t$ decides the step size of gradient descent.

% \begin{table*}
% \caption{Budget Allocation Schedules($\sigma_0=8$)}
% \centering
% \begin{tabular}{|c|c|c|c|}
% \hline
% Time-Based Decay& Step Decay& Exponential Decay & Polynomial Decay\\
% k=0.016& k=0.78, period=20 & k=0.01 & k=2, $\sigma_{end}=1$
% \\
% \hline
% \includegraphics[scale=0.3]{./figures/e1}&\includegraphics[scale=0.3]{./figures/e2} &
% \includegraphics[scale=0.3]{./figures/e3}&\includegraphics[scale=0.3]{./figures/e4}\\
% \hline
% \end{tabular}
% \label{tab:gt}
% \end{table*}

\subsubsection{Privacy Preserving Parameter Selection}
The proposed schedules require a set of pre-defined hyperparameters, such as decay rate and period. Their values decide the training time and affect the final model accuracy. It is expected to find the optimal hyperparameters for the schedules to produce the most accurate model. A straightforward approach is to test a list of $k$ candidates by training $k$ neural networks respectively and trivially choose the one that achieves the highest accuracy, though this adds the privacy cost up to $k \rho_{total}$. A better approach is to apply differentially private parameter tuning, such as the mechanism proposed by Gupta et al.~\cite{Chaudhuri:2011:DPE}. The idea is to partition the dataset to $k+1$ equal portions, train $k$ models with using $k$ schedules on $k$ different data portions respectively, and evaluate the number of incorrect predictions for each model, denoted by $z_i$ ($1\le i\le k$), on the remaining data portion. Then, the Exponential Mechanism~\cite{McSherry:2007:MDV} is applied, which selects and outputs a candidate with the probability proportional to $exp(\frac{-\epsilon z_i }{2})$. This parameter tuning procedure satisfies $\epsilon$-DP, and accordingly satisfies $\frac{1}{2}\epsilon^2$-zCDP~\cite{BunS16}.

\subsection{Refined Privacy Accountant}
\label{sec:privacyacccount}
The composition property of zCDP allows us to easily compute cumulative privacy loss for the iterative SGD training algorithm. Suppose that each iteration satisfies $\rho$-zCDP and the training runs $T$ iterations, then the whole training process satisfies ($T\rho$)-zCDP.
In this section we show that 1) the composition can be further refined by considering the property of the mini-batch SGD algorithm, and 2) more importantly, different batching methods lead to different privacy loss. In particular, we analyze the privacy loss composition under two common batching methods: random sampling with replacement and random reshuffling.
With random reshuffling, the training dataset is randomly shuffled and then partitioned into batches of similar size and SGD sequentially processes one batch at a time. It is a random sampling process without replacement. For random sampling with replacement, each example in a batch is independently sampled from the training dataset with replacement.
Because these two data batching methods have different privacy guarantees, for tracking privacy loss correctly, it is important for the users to choose the right accounting method based on the batching method they use.  

\subsubsection{Under random reshuffling}
\label{ssec:privacyRF}
SGD takes disjoint data batches as input within an epoch with random reshuffling. We note that the existing results~\cite{Dwork2016ConcentratedPrivacy,BunS16,Dwork:2006:CNS} on the composition of a sequence of differential private mechanisms $\mathcal{A}_1, \ldots, \mathcal{A}_k$ assume that each mechanism $\mathcal{A}_i$ runs with the same dataset $X$ as input. It is expected that their composition has less cumulative privacy loss if each of differentially private mechanisms runs on disjoint datasets. The formal composition result in this scenario is detailed in Theorem~\ref{recompose}. All the proofs of lemmas and theorems in this paper can be found in Appendix.

\begin{theorem}
\label{recompose}
Suppose that a mechanism $\mathcal{A}$ consists of a sequence of $k$ adaptive mechanisms, $\mathcal{A}_1,\ldots, \mathcal{A}_k$, where each $\mathcal{A}_i:~ \prod_{j=1}^{i-1}\mathcal{R}_j \times \mathcal{D} \rightarrow \mathcal{R}_i$ and $\mathcal{A}_i$ satisfies $\rho_i$-zCDP ($1 \leq i \leq k$). Let $\mathbb{D}_1, \mathbb{D}_2, \ldots, \mathbb{D}_k$ be the result of a randomized partitioning of the input domain $\mathbb{D}$.
The mechanism $\mathcal{A}(D) = (\mathcal{A}_1(D\cap \mathbb{D}_1), \ldots,\mathcal{A}_k(D\cap \mathbb{D}_k))$ satisfies
$\max\limits_i \rho_i$-zCDP.
\end{theorem}

\begin{corollary}
\label{recomposesamerho}
Suppose that a mechanism $\mathcal{A}$ consists of a sequence of $k$ adaptive mechanisms, $\mathcal{A}_1,\ldots, \mathcal{A}_k$, where each $\mathcal{A}_i:~ \prod_{j=1}^{i-1}\mathcal{R}_j \times \mathcal{D} \rightarrow \mathcal{R}_i$ and $\mathcal{A}_i$ satisfies $\rho$-zCDP ($1 \leq i \leq k$). Let $\mathbb{D}_1, \mathbb{D}_2, \ldots, \mathbb{D}_k$ be the result of a randomized partitioning of the input domain $\mathbb{D}$.
The mechanism $\mathcal{A}(D) = (\mathcal{A}_1(D\cap \mathbb{D}_1), \ldots,\mathcal{A}_k(D\cap \mathbb{D}_k))$ satisfies
$\rho$-zCDP.
\end{corollary}

Theorem \ref{recompose} provides a tighter characterization of privacy loss for the composition of mechanisms having disjoint input data.  
Assume for all $i \in (1,\ldots, k)$, $\rho_i= \rho$. Given Theorem~\ref{recompose}, it is trivial to show that mechanism $\mathcal{A}$ satisfies $\rho$-zCDP, stated in Corollary \ref{recomposesamerho}. Compared with a guarantee of $(k \rho)$-zCDP using the composition property in Proposition \ref{prop:Gaussianzcdp}, 
we demonstrate that the total privacy loss of sequential computations on disjoint datasets is just $\rho$, equivalent with one computation step in the sequence.
For $\epsilon$-DP, the similar result in~\cite{McSherry:2009:PIQ} of the parallel composition theorem says that when each $\epsilon_i$-differentially private mechanism queries a disjoint subset of data in parallel and work independently, their composition provides $(\max_i \epsilon_i)$-DP instead of the $\sum_i \epsilon_i$ derived from a naive composition. 
   
In the differentially private mini-batch SGD algorithm shown in Algorithm~\ref{alg:dpsgd}, each iteration step $t$ satisfies $(\frac{1}{2 \sigma_t^2})$-zCDP according to Proposition \ref{prop2:Gaussianzcdp}. In this paper, we suppose that each iteration in the same epoch uses the same noise scale $\sigma$ and each uses a disjoint data batch. Then, by Theorem \ref{recompose}, we know that the computation of this epoch still satisfies ($\frac{1}{2 \sigma^2}$)-zCDP. Because the training dataset is repeatedly used every epoch, the composition of the epoch level computations follows normal composition of Proposition \ref{prop:Gaussianzcdp}. Thus, when the training runs a total of $E$ epochs and each epoch $e$ satisfies $\rho_e$-zCDP, the whole training procedure satisfies ($\sum_{e=1}^E \rho_e$)-zCDP.

\noindent\textbf{Note}: In our previous conference version the statement of Theorem 2 that uses an average $\frac{1}{k} \sum \rho_i$-zCDP only holds when all $\rho_i$ are the same, i.e., $\rho_i=\rho$. Generally, it should be replaced with the max as stated in the revised Theorem \ref{recompose} of this paper. Because our algorithm and experiments in~\cite{Yu2018DifferentiallyPM} assume the same noise scale in each iteration within the same epoch, i.e., the same $\rho_i$, our analysis and experiment results remain correct.

\subsubsection{Under random sampling with replacement}
\label{ssec:sampling}
We have shown a privacy amplification effect resulting from the disjoint data access of every iteration within one epoch under random reshuffling. In contrast, the MA method~\cite{Abadi:2016:DLD} exploits the privacy amplification effect of random sampling with replacement. In this section, we examine how random sampling with replacement affects the privacy loss in terms of zCDP.
Intuitively, the random sampling with replacement introduces more uncertainty than the random reshuffling process which samples data batches without replacement. However, our analysis shows that CDP cannot characterize the privacy amplification effect of random sampling. It is because of the restrictive notion of sub-Gaussianity in CDP which requires moment constraints on all orders, i.e., $\alpha \in (1, \infty)$ in Eq. (\ref{eq:zcdpeq}). To address this problem, we propose a relaxation of zCDP and convert it to $(\epsilon, \delta)$-DP. This then allows us to capture the privacy amplification of random sampling.

Suppose a new mechanism $\mathcal{A'}$ that runs $\rho$-zCDP mechanism $\mathcal{A}$ on a random subsample of dataset $\mathbb{D}$ where each example is independently sampled with probability $q$.
Without loss of generality, we fix $\mathbb{D}$ and consider a neighboring dataset $\mathbb{D}'$= $\mathbb{D} \cup {d_e}$.
we use $\Lambda(*)$ to denote the sampling process over dataset $*$, let $T$ be any subsample that does not include $d_e$ and $T'=T \cup{d_e}$. Because $d_e$ is randomly sampled with probability $q$, $\mathcal{A'(\mathbb{D}')}$ is distributed identically as 
$u_0\overset{\Delta}{=} \Pr(\Lambda(\mathbb{D})=T)\mathcal{A}(T)$ with probability $(1-q)$, and as $u_1\overset{\Delta}{=} \Pr(\Lambda(\mathbb{D}')=T'|d_e \in T') \mathcal{A}(T')$ with probability $q$.
Therefore, the following holds:
\begin{equation}
\label{eq:stillrho}
\mathcal{A'}(\mathbb{D}) \sim u_0 ,~~~~~~    \mathcal{A'}(\mathbb{D}') \sim q u_1 +(1-q) u_0
\end{equation}
By $\Pr\big(\Lambda(\mathbb{D})=T\big)$=$\Pr\big(\Lambda(\mathbb{D}')=T'|d_e \in T'\big)$ due to $T'=T \cup{d_e}$, it is easy to prove that $\mathcal{A'}$ still satisfies $\rho$-zCDP. The proof details can be found in the Appendix \ref{ssec:samerho}. 

\noindent \textbf{No privacy amplification for $\mathcal{A'}$ in terms of CDP}.
Consider
\begin{align}
&(\alpha-1)D_\alpha {(\mathcal{A'}(\mathbb{D}')||\mathcal{A'}(\mathbb{D}))}  \notag \\
&=\log \bigg(\mathbb{E}_{u_0}\left[ \big(\frac{q u_1 +(1-q) u_0}{u_0}\big)^\alpha \right] \bigg)
>\log \bigg(\mathbb{E}_{u_0}\left[ \big(q\frac{u_1}{u_0}\big)^\alpha \right] \bigg) \notag \\
&=\alpha \log q + (\alpha-1)D_\alpha(u_1||u_0) \label{a2}
\end{align}
When $\alpha \rightarrow \infty$, we have 
\begin{equation}
D_\alpha {(\mathcal{A'}(\mathbb{D}')||\mathcal{A'}(\mathbb{D}))} > \alpha \rho + \frac{\alpha}{\alpha-1}\log q \rightarrow \alpha \rho + \log q
\end{equation}
This shows that the sampling does not produce any reduction with regard to $q$ on $\rho$ and still $D_\alpha {(\mathcal{A'}(\mathbb{D}')||\mathcal{A'}(\mathbb{D}))}= \Theta(\alpha \rho)$. Therefore, by definition, zCDP is not able to capture the privacy amplification effect of random sampling.

\begin{figure}
    \centering
    \includegraphics[width=0.25\textwidth]{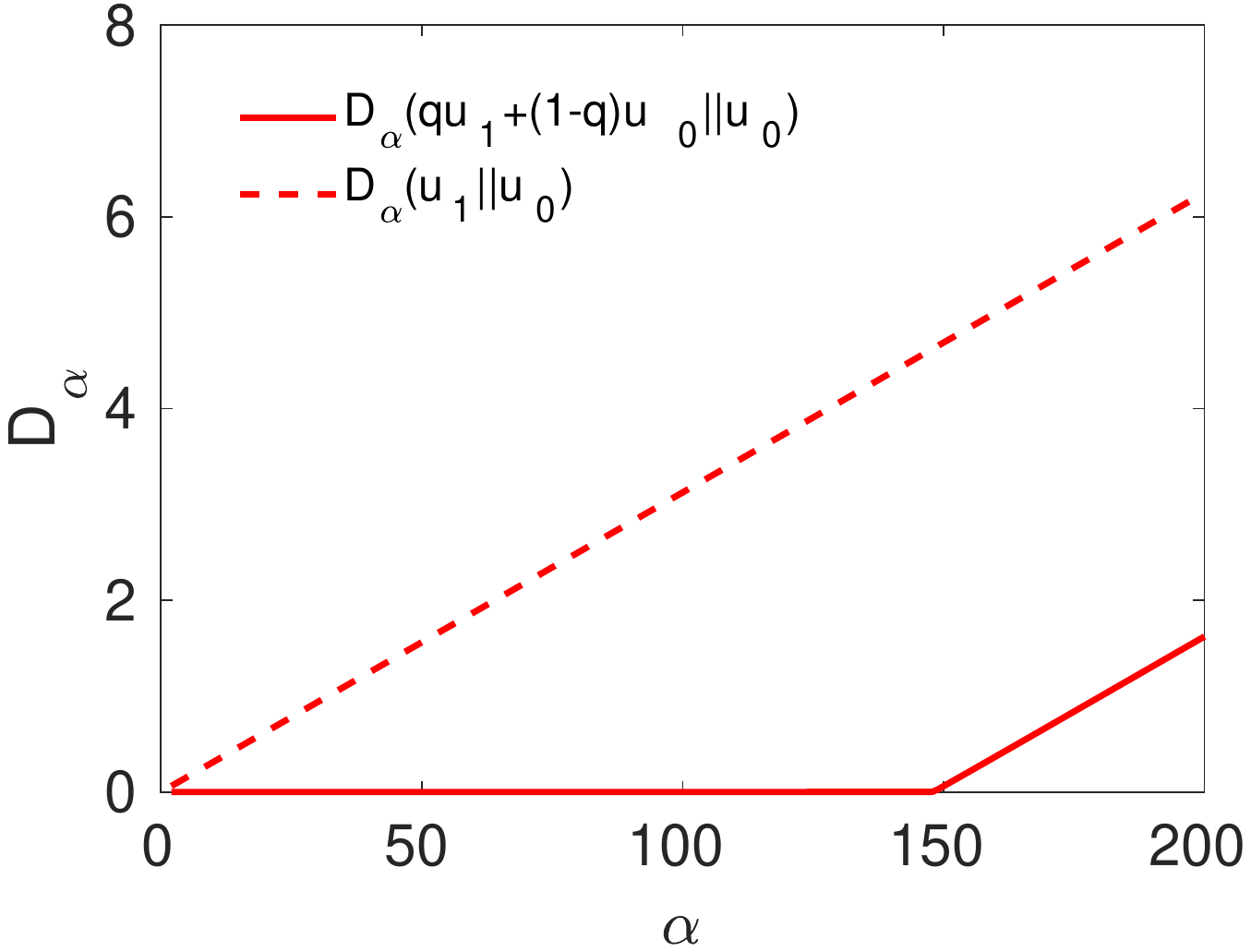}
    \caption{$\alpha$-R\'enyi divergence under sampling($q$=0.01, $\sigma$=4)}
    \label{fig:moments}
\vspace{-12pt}
\end{figure}

The reason there is no privacy amplification with random sampling on zCDP is that concentrated DP requires a sub-Gaussian distribution for privacy loss, and thus moments must be bounded by $\exp(O(\alpha^2))$ at all orders $\alpha \in (1, +\infty)$. This is a fairly strong condition. Alternatively, it requires that a mechanism $\mathcal{A}$ has $\alpha$-R\'enyi divergence bounded by $\alpha \rho$ for all $\alpha \in (1, +\infty)$. To demonstrate that, we assume a Gaussian mechanism $\mathcal{A}$ that $\mathcal{A}(T)\sim N(0,\sigma^2)$ and $\mathcal{A}(T')\sim N(1,\sigma^2)$. Denote $N(0,\sigma^2)$ by $p_0$ and $N(1,\sigma^2)$ by $p_1$. It is trivial to show that
$D_\alpha(u_1||u_0) = D_\alpha(p_1||p_0)$ and $D_\alpha(q u_1 +(1-q) u_0||u_0) = D_\alpha(q p_1 +(1-q) p_0||p_0)$.
We can then numerically compute $D_\alpha(u_1||u_0)$ and $D_\alpha(q u_1 +(1-q) u_0||u_0)$ with $q=0.01$ and $\sigma=4$. Results are shown in Figure \ref{fig:moments}. $D_\alpha(u_1||u_0)$ is linear in $\alpha$ because $D_\alpha(u_1||u_0)=\alpha \rho= \alpha / (2\sigma^2)$. However, we can see that $D_\alpha(q u_1 +(1-q) u_0||u_0)$ has a changing point at $\alpha=147$. Before this changing point it is close to zero, because $q=0.01$ is very small and the two distributions $q u_1 +(1-q) u_0$ and $u_0$ are close to each other. At higher orders, the sampling effect vanishes and the divergence increases at the rate of $D_\alpha(u_1||u_0)$. This indicates that the privacy amplification effect from random sampling does not hold for all orders $\alpha \in (1, +\infty)$, and we cannot improve the privacy metric in terms of zCDP under the random sampling. On the other hand, Figure~\ref{fig:moments} suggests that we need to analyze the $\alpha$-R\'enyi divergence within a limited range of $\alpha$ to capture the privacy amplification effect.
We will show that having a bound on $D_\alpha {(\cdot||\cdot)}$ within a limited range of $\alpha$ makes it possible to capture the privacy amplification effect of random sampling. However, such constraint does not fit into the definition of CDP since it indicates a sub-exponential privacy loss variable, a relaxation to sub-Gaussianity in the definition of CDP. Therefore, in this paper we address it by converting the $\alpha$-R\'enyi divergence under such relaxation to traditional ($\epsilon$, $\delta$)-DP. 

In the following we show the conversion to $(\epsilon,\delta)$-DP in a general form for a Gaussian mechanism with bounded $\alpha$-R\'enyi divergence in some limited range of $\alpha$ parameterized by $q$ and $\sigma$,  and then consider specific cases with concrete values for the range of $\alpha$ and bounds on $\alpha$-R\'enyi divergence.

Suppose that $f: \mathbb{D} \rightarrow \mathbb{R}^p$ with $||f(\cdot)||_2 \le 1$. Consider a mechanism $\mathcal{A'}$ that runs a Gaussian mechanism adding noise $\mathcal{N}(0, \sigma^2 \mathbb{I})$ over a random subsample $J \subseteq \mathbb{D}$ where each example is independently sampled with probability $q$, i.e., $\mathcal{A'}(\mathbb{D}) = \sum_{i \in J} f(x_i) + \mathcal{N}(0, \sigma^2 \mathbb{I})$. Let $D_\alpha {(\cdot||\cdot)}$ be the $\alpha$-R\'enyi divergence between $\mathcal{A'}(\mathbb{D})$ and $\mathcal{A'}(\mathbb{D}')$ for two neighboring datasets $\mathbb{D}$ and $\mathbb{D}'$. Consider $P(*)$ and $U_\alpha(*)$ as some finite functions of $q$ and $\sigma$. We assume $D_\alpha {(\cdot||\cdot)}$ is bounded by $\alpha \cdot P(q, \sigma)$ within a limited range of $1 < \alpha \le U_\alpha(q, \sigma)$. Then, we have the following theorem: 

\begin{theorem}
\label{recompose3}
Let $\widehat{\rho} =P(q,\sigma)$ and $u_\alpha = U_\alpha(q, \sigma)$.
If the mechanism $\mathcal{A'}$ has
\begin{equation}
\label{th:1}
D_\alpha (\mathcal{A'}(\mathbb{D})||\mathcal{A'}(\mathbb{D}')) \le \alpha \widehat{\rho}
\end{equation}
for $1<\alpha \le u_\alpha$, it satisfies
\begin{numcases}{}
\small
\hspace{-5pt}\Big(\widehat{\rho} + 2\sqrt{\widehat{\rho}\log(1/\delta)},\delta \Big)-DP, \text{~if~} \delta \ge  1/\exp(\widehat{\rho}(u_\alpha-1)^2) &  \label{eq:dpcases1} \\
\hspace{-5pt}\Big(\widehat{\rho}u_\alpha-\frac{\log \delta}{u_\alpha-1},\delta \Big)-DP, \text{otherwise} &  \label{eq:dpcases2}
\end{numcases}
\end{theorem}

Theorem \ref{recompose3} shows in a general form the connection to ($\epsilon$, $\delta$)-DP when we have $\alpha$-R\'enyi divergence bounded within a limited range of $\alpha$. Proper choice of $U_\alpha(q,\sigma)$ for the range of $\alpha$ (for example, let $U_\alpha(q,\sigma) < 147$ in Figure 1) with a corresponding $P(q, \sigma)$ can capture the privacy amplification effect of random sampling with replacement, which we will discuss later.

%\begin{lemma}
%\label{recompose2}
%Suppose that $f: \mathbb{D} \rightarrow \mathbb{R}^p$ with $||f(\cdot)||_2 \le 1$.
%Consider a mechanism $\mathcal{A'}$ that runs a Gaussian mechanism adding noise $\mathcal{N}(0, \sigma^2 \mathbb{I})$ over a random subsample $J \subseteq \mathbb{D}$ where each example is independently sampled with probability $q$, i.e., $\mathcal{A'}(\mathbb{D}) = \sum_{i \in J} f(x_i) + \mathcal{N}(0, \sigma^2 \mathbb{I})$. For two neighboring datasets $\mathbb{D}$ and $\mathbb{D}'$,
%the $\alpha$-R\'enyi divergence between  $\mathcal{A'}(\mathbb{D})$ and $\mathcal{A'}(\mathbb{D}')$ is given below:
%\begin{equation}
%\begin{split}
%D_\alpha {(\cdot||\cdot)} \le \frac{q^2\alpha}{(1-q)\sigma^2}+O(q^3(\alpha-1)^2/\sigma^3), \\ \rm{~~~for~~} 1<\alpha \le \sigma^2 \log \frac{1}{q \sigma}+1
%\end{split}
%\label{th:1}
%\end{equation}
%\end{lemma}

\noindent\textbf{Composition.}~
Now we consider the composition of a sequence of Gaussian mechanisms with random sampling. Suppose $k$ mechanisms, denoted by $\mathcal{M}$=$(\mathcal{A}'_1$,\ldots, $\mathcal{A}'_k)$ where each $\mathcal{A}'_i$ uses sampling ratio $q_i$ and noise scale $\sigma_i$. 
Because the constraint of $\alpha$ in Eq. (\ref{th:1}) depends on the sampling ratio and noise scale, we examine their composition in two cases: 

1.) Each mechanism uses the same $q$ and $\sigma$. For $1<\alpha \le u_\alpha$,  by the composition property of $\alpha$-R\'enyi divergence~\cite{BunS16}, we have
$D_\alpha (M(\mathbb{D})||M(\mathbb{D}')) \le k\alpha P(q,\sigma)$. The conversion to $(\epsilon,\delta)$-DP can be done by letting $\widehat{\rho} = k P(q,\sigma)$ in (\ref{eq:dpcases1}) and (\ref{eq:dpcases2}) in Theorem \ref{recompose3}.

2.) The sampling ratio and noise scale are different for each mechanism. Then, for each $\mathcal{A}_i$, we have $D_\alpha (\mathcal{A}_i(\mathbb{D})||\mathcal{A}_i(\mathbb{D}')) \le \alpha P(q_i,  \sigma_i)$ for $1<\alpha \le U_\alpha(q_i,  \sigma_i)$. To allow the composition of $\alpha$-R\'enyi divergence of mechanisms with different $q_i$ and $\sigma_i$, we constrain $\alpha$ to the range $1< \alpha \le \min_i \{U_\alpha(q_i,  \sigma_i)\}$. It is then clear that
$D_\alpha (M(\mathbb{D})||M(\mathbb{D}')) \le \alpha (\sum_i P(q_i, \sigma_i))$ holds within this $\alpha$ range.
Letting $\widehat{\rho}=\sum_i P(q_i, \sigma_i)$ and replacing $q$ and $\sigma$ by $q_j$ and $\sigma_j$ where $j=\argmin _i \{U_\alpha(q_i,  \sigma_i)|1 \le i\le k\}$ in Eq. (\ref{eq:dpcases1}) and (\ref{eq:dpcases2}), we can still obtain the corresponding ($\epsilon$,$\delta$)-DP.

When random sampling is used for batching, Algorithm \ref{alg:dpsgd} follows the above method to estimate privacy loss in terms of $(\epsilon,\delta)$-DP. In particular, the algorithm specifies a fixed $\delta = \delta_0$ and a total privacy budget $\epsilon_{total}$, and at every iteration step $t$, it updates $\widehat{\rho} = \sum_{i=0}^t P(q_i, \sigma_i)$ and computes the corresponding cumulative privacy loss $\epsilon_t$. If $\epsilon_t > \epsilon_{total}$, the training is terminated and the final model satisfies $(\epsilon_{total}, \delta_0)$-DP. 

\noindent\textbf{The empirical settings of $P$ and $U_\alpha$.} To put Theorem \ref{recompose3} and its composition in use to estimate privacy loss while capturing the privacy amplification effect of random sampling, proper $U_\alpha(q, \sigma)$ and $P(q,\sigma)$ have to be determined. However, the $\alpha$-R\'enyi divergence for Gaussian mechanism with sampling appears to be analytically intractable. We noted an asymptotic bound on the log moment was given in previous work~\cite{Abadi:2016:DLD} when $q \le \frac{1}{16\sigma}$. By definition, $\alpha$-R\'enyi divergence is equal to the log moment multiplied by a factor of $\frac{1}{\alpha-1}$. Then, under the same condition, $D_\alpha {(\cdot||\cdot)}$ has an asymptotic bound of $\frac{q^2}{1-q} \alpha / \sigma^2 + O(q^3(\alpha-1)^2/\sigma^3)$ for $1<\alpha \le \sigma^2 \log \frac{1}{q \sigma}+1$. This bound exhibits the privacy amplification by having a factor of $q^2$. 
Therefore, our solution here is to empirically find an appropriate $P(q, \sigma)$ around the asymptotic bound within $1<\alpha \le$ $U_\alpha(q,\sigma)=\sigma^2 \log \frac{1}{q \sigma}+1$ given $q \le \frac{1}{16\sigma}$.
 
We set $P(q, \sigma)=q^2/\sigma^2$ which is originally derived with the log moment bound given in the proof statement of Theorem 1 of \cite{Abadi:2016:DLD}. It was brought to our attention that this bound misses an asymptotic term. However, by conducting numerical comparison, we observed that with using $P(q, \sigma)=q^2/\sigma^2$, the bound $\alpha q^2/\sigma^2$ for $\alpha$-R\'enyi divergence in (\ref{th:1}) empirically holds for a wide of parameter settings of $q$ and $\sigma$. Our numerical validations on this bound and the details are given in Appendix \ref{Appendix:v}. Therefore, this approach enables a quick estimation of privacy loss which can be easily used to decide appropriate noise scale, sampling ratio and the number of training steps during the planning phase. especially when we choose dynamic privacy budget allocation. It is worth to note, for privacy accounting with a provable guarantee in a more general setting and tighter results, the MA method~\cite{Abadi:2016:DLD} should be used.
 
\noindent\textbf{Summary.} We have shown that CDP is not able to capture the privacy amplification effect of random sampling. We consider bounding $\alpha$-R\'enyi divergence over a constrained range of $\alpha$ instead of $(1, \infty)$ and convert to $(\epsilon,\delta)$-DP. Our empirical solution can be used as an approximate privacy loss estimation and easy to compute especially when we use different sampling ratios and noise scales for each iteration of DP-SGD for dynamic privacy budget allocation.

More importantly, we have provided formal analysis to show that the compositions of differential privacy under two batching methods are distinct. As demonstrated by our experimental results, this causes different privacy loss. Therefore, we argue that the privacy accounting method has to be chosen according to which data batching method is used. In our implementation, we focus on random reshuffling, because it is a common practice~\cite{inputpipe,Goodfellow-et-al-2016} followed by existing deep learning implementations and it is also numerically observed that random reshuffling outperforms random sampling with replacement in the convergence rate~\cite{2015arXivG}.

\noindent\textbf{Note}: In our previous conference version we set $P(q, \sigma)=q^2/\sigma^2$  directly based on the derivation from the log moment bound stated in the proof of Theorem 1 of \cite{Abadi:2016:DLD} which misses an asymptotic term. Our experiment numerically validates this bound ready to use in a wide range of parameter settings.

\subsection{DP Composition Under Dynamic Schedules}
For pre-defined schedules, once the hyperparameters are specified, they follow the decay functions to update the noise scale without accessing the data and the model, and thus do not incur any additional privacy cost.
Since the noise scale is updated by epoch, each iteration step within an epoch uses the same noise scale. Suppose the epoch $t$ uses the noise scale $\sigma_t$. Each iteration of epoch $t$ is then $\rho=1/(2\sigma_t^2)$ zCDP by Proposition \ref{prop2:Gaussianzcdp}, and the total privacy cost of epoch $t$ can be calculated by Theorem \ref{recompose} or \ref{recompose3} depending on which batching method is used. Over the course of training, the cumulative privacy loss is updated at each epoch, and once the cumulative privacy loss exceeds the fixed privacy budget $\rho_{total}$, the training is terminated.
To achieve a target training time under a given total privacy budget, we can determine the exact values of hyperparameters for these schedules in advance before the training time.

For the validation-based schedule, the access to the public validation dataset does not incur additional privacy cost. With this schedule, the composition of differential privacy involves adaptive choices of the privacy parameter $\rho$ at every epoch, which is corresponding to the noise scale $\sigma$ of the Gaussian mechanism. This means that the choice of privacy parameters itself is a function of the realized outcomes of the previous rounds. It has been shown by Rogers et al.~\cite{conf/nips/RogersVRU16} that the strong composition theorem for $(\epsilon,\delta)$-DP fails to hold in this adaptive privacy parameter setting since the theorem requires the privacy parameters to be pre-defined ahead of time. To address this problem, they define the privacy loss as a random variable as done in Eq. (\ref{eq:privacyloss}) for CDP and develop the composition for $(\epsilon,\delta)$-DP using privacy filters. Privacy filters provide a way to halt the computation with probability $1-\delta$ before the realized privacy loss exceeds $\epsilon$.
Our approach relies on zCDP which defines privacy loss as in Eq. (\ref{eq:privacyloss}) by nature and therefore the composition accumulating privacy cost with regard to R\'enyi divergence holds for the adaptive parameter settings.

\section{Experimental Results}
\label{sec:exp}
In this section, we evaluate the proposed privacy accounting methods, and demonstrate the effectiveness of dynamic privacy budget allocation on different learning tasks. Our implementation is based on the TensorFlow implementation~\cite{tfdpcode} of DP-SGD in the paper~\cite{Abadi:2016:DLD}.

\subsection{Comparing Privacy Accounting Approaches}
\label{ssec:pa}
In Section \ref{sec:privacyacccount} we derive different privacy accounting methods for two data batching methods: random reshuffling (RF) and random sampling with replacement (RS). We refer to them as zCDP(RF) and zCDP(RS) respectively. To numerically compare them with other privacy accounting methods including strong composition~\cite{advcomposition} and the moments accountant (MA) method~\cite{Abadi:2016:DLD}, we
unify them into  $(\epsilon, \delta)$-DP. 
Following~\cite{Abadi:2016:DLD}, we assume that the batches are generated with RS for both the strong composition and MA. We use the implementation of \cite{Abadi:2016:DLD} in TensorFlow to compute MA. For strong composition, we apply the strong composition theorem in~\cite{advcomposition} to the composition of $(\log(1+q(e^\epsilon-1)), q\delta)$-DP mechanisms that are the privacy amplified version of $(\epsilon, \delta)$-DP mechanisms running with random sampling ratio $q$.
We compute $(\epsilon, \delta)$-DP for zCDP(RF) with Proposition \ref{prop:zcdpvsdp} and for zCDP(RS) with the methods in Section \ref{ssec:sampling}.

In our experimental setting, we assume a batch size $B$ for random reshuffling. For random sampling we assume a sampling ratio $q=\frac{B}{N}$ given a total of $N$ samples. In the following, when we vary $q$, it is equivalent to change the batch size for random reshuffling.
The number of iterations in one epoch is $\frac{1}{q}$.  
For simplicity, we use the same noise scale for Gaussian mechanism $\mathcal{N}(0, C^2 \sigma^2 \mathbf{I})$ for every iteration and set $q=0.01$ and $\sigma=6$ by default. Given $\sigma=6$, the Gaussian mechanism satisfies both ($\epsilon=0.808$, $\delta=1e-5$)-DP and $\rho=0.0139$-zCDP.
We track the cumulative privacy loss by epoch with different privacy accounting methods and convert the results to $\epsilon$ in terms of ($\epsilon, \delta)$-DP with fixed $\delta=1e-5$.

\begin{figure}[t]
    \centering
    \includegraphics[width=0.27\textwidth, height=0.2\textwidth]{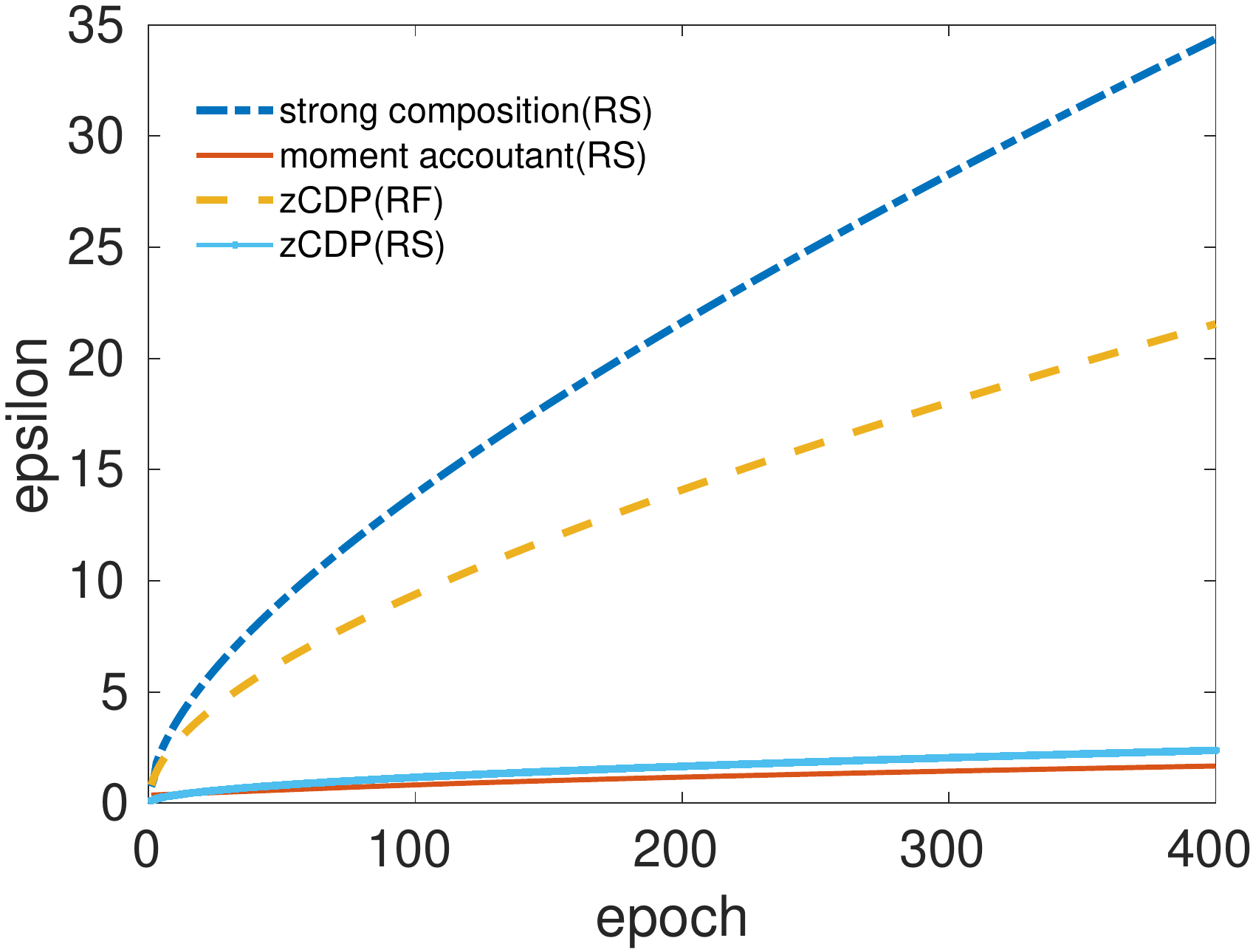}
    \caption{Privacy parameter $\epsilon$ v.s. epoch}
    \label{fig:accoutant}
\centering
\begin{subfigure}{0.23\textwidth}
\includegraphics[width=0.9\textwidth, height=0.7\textwidth]{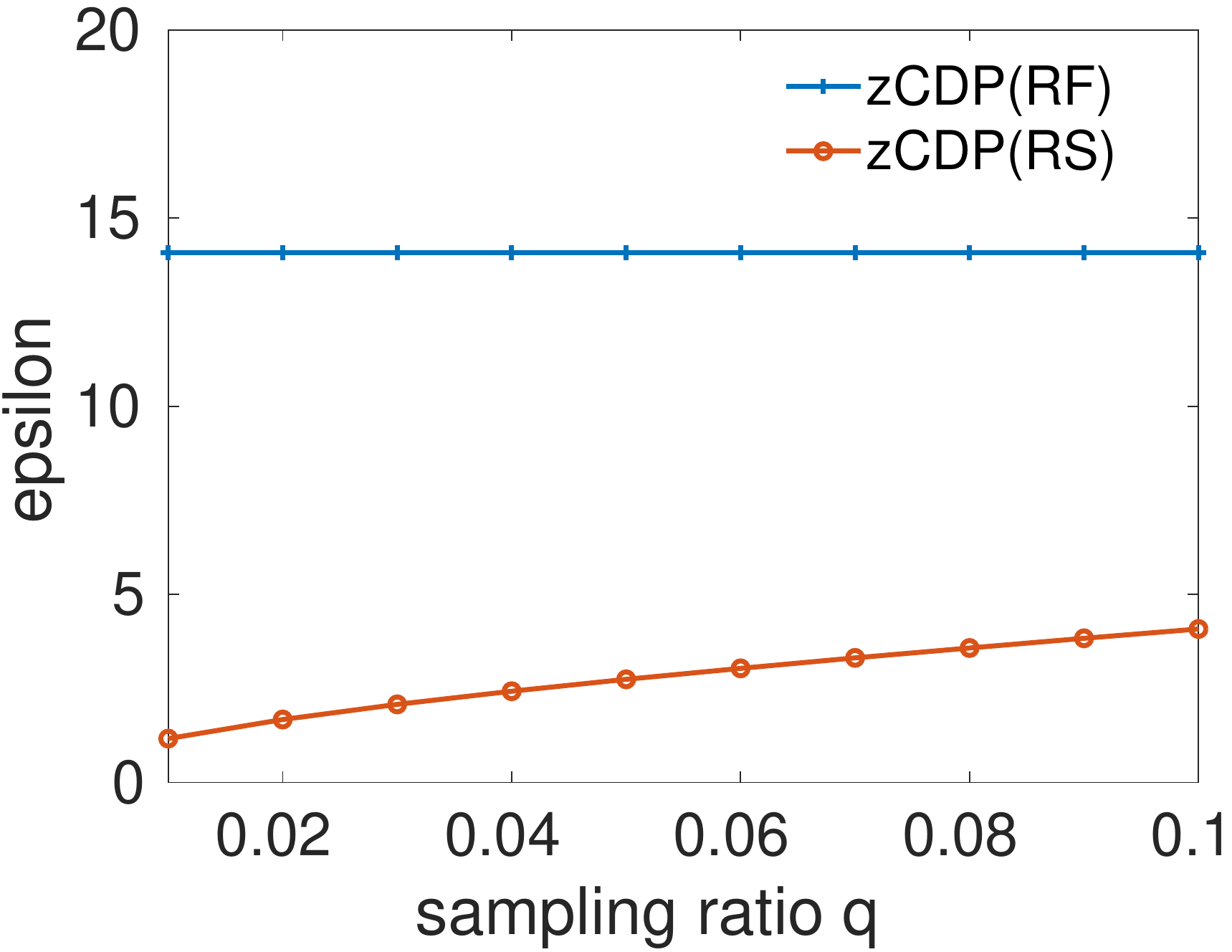}
    \caption{ $\epsilon$ v.s. $q$}
    \label{fig:samplingratio}
\end{subfigure}
\begin{subfigure}{0.23\textwidth}
    \centering
    \includegraphics[width=0.9\textwidth,height=0.68\textwidth]{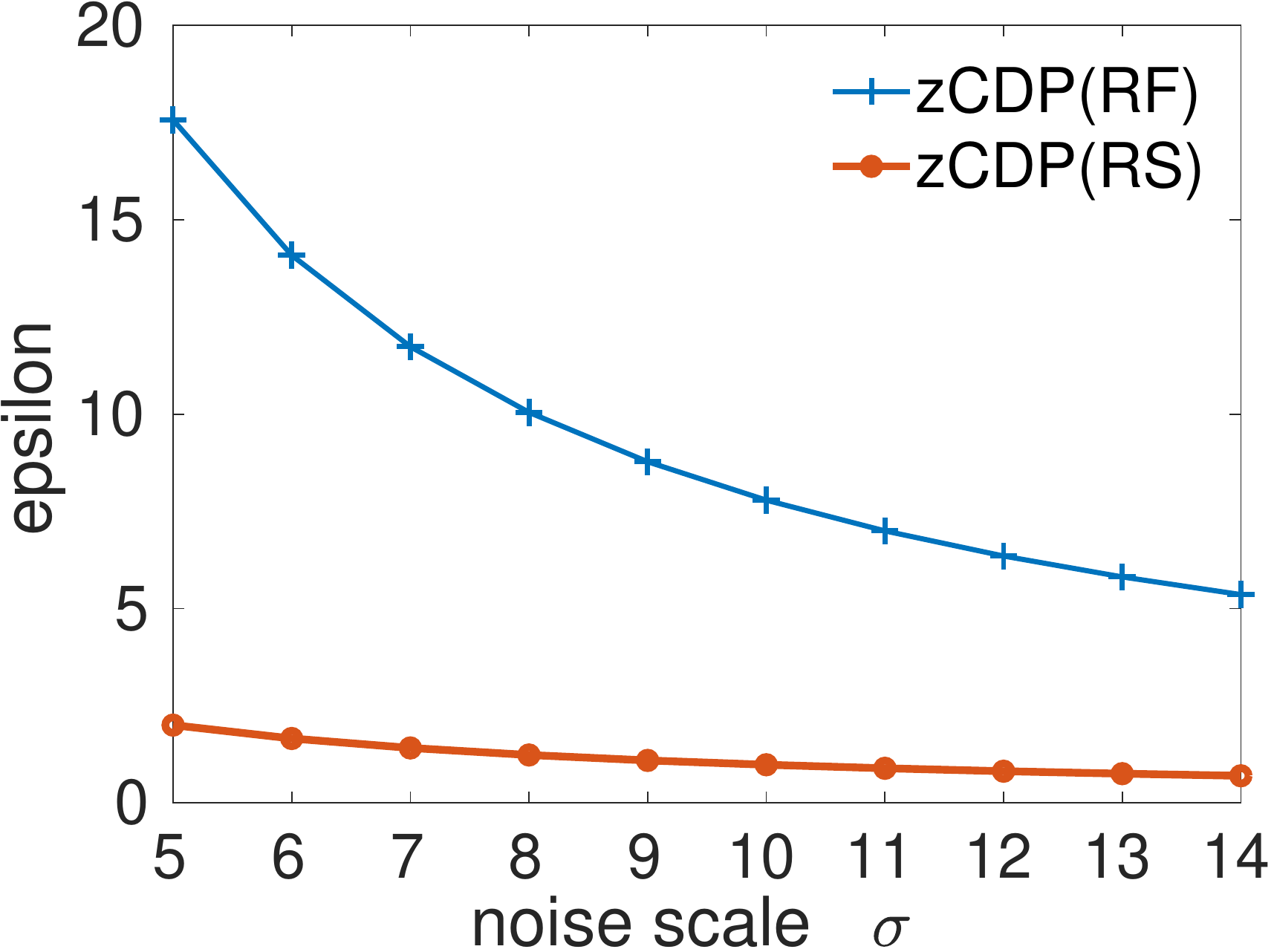}
    \caption{ $\epsilon$ v.s. $\sigma$}
    \label{fig:sigma}
\end{subfigure}
\caption{privacy loss $\epsilon$ v.s. sampling ratio $q$ \& noise scale $\sigma$}
\vspace{-16pt}
\end{figure}

Figure \ref{fig:accoutant} shows the growth of privacy loss metric $\epsilon$ during the training process.
It shows that zCDP(RF) has lower estimation on privacy loss than that of the strong composition during the training. The final spent $\epsilon$ at epoch 400 by zCDP(RF) and strong composition are 21.5 and 34.3 respectively. Although random sampling introduces higher uncertainty and thus less privacy loss than random shuffling, zCDP(RF) still achieves lower and thus tighter privacy loss estimation even than the strong composition with random sampling. This demonstrates the benefit of CDP for composition of a large number of computations.
The results for zCDP(RS) and MA are very close to each other because that they both exploit the moment bounds of privacy loss to achieve tighter tail bound and take advantage of the privacy amplification of random sampling. The final spent $\epsilon$ is 2.37 and 1.67 for zCDP(RS) and MA respectively. The reason for zCDP(RS) to have a slightly higher estimation is that its conversion to $(\epsilon, \delta)$-DP explicitly uses an upper bound for the log moment instead of the numerical computation of log moments. The benefit of zCDP(RS) is that it is simple to compute with explicit expressions in Theorem \ref{recompose3}, and its composition is easy under dynamic privacy budget allocation.

Figure \ref{fig:accoutant} shows that zCDP(RF) has higher privacy loss compared to MA and zCDP(RS), because more uncertainty is introduced with RS. However, it is worth noting that the common practice in deep learning is to use RF, including the implementation of \cite{Abadi:2016:DLD}.
Thus, zCDP(RF) is the proper choice for them and also straightforward due to the composition property of $\rho$-zCDP which simply adds up on $\rho$ values. The results show that MA underestimates the real privacy loss when treating the random reshuffling as random sampling with replacement.

We further examine how zCDP(RF) and zCDP(RS) change with the sampling ratio $q$ and the noise scale $\sigma$. Using the default $\sigma=6$, Figure \ref{fig:samplingratio} shows the privacy loss $\epsilon$ at the end of 200 training epochs with varying $q$ values. For zCDP(RF), the cumulative privacy loss does not change with $q$. This is because the composition of $\rho$-zCDP iterations within one epoch still satisfies $\rho$-zCDP by Theorem \ref{recompose} and across epochs the linear composition of $\rho$-zCDP in Theorem \ref{prop:Gaussianzcdp} is applied, which makes the final privacy loss depend exclusively on the number of training epochs. We have fixed 200 epochs so the final privacy loss does not change.
In contrast, the privacy loss given by zCDP(RS) increases with the sampling ratio $q$, which can be seen in Eq. (\ref{eq:dpcases1}) where $\epsilon$ increases with $\widehat{\rho}$ which is proportional to $q^2$.
Similarly, Figure \ref{fig:sigma} shows the privacy loss after 200 epochs by varying noise scales with the same $q$=$0.01$. We observe that increasing $\sigma$ from 5 to 14 significantly reduces $\epsilon$ for zCDP(RF) but has noticeably less impact on zCDP(RS). It suggests that under random sampling, a small sampling ratio contributes much more on privacy than the noise scale $\sigma$. This indicates that we may reduce the noise scale to improve the model accuracy without degrading much privacy. However, for random reshuffling, the privacy loss does not depend on the sampling ratio (i.e., the batch size) but is decided by $\sigma$, so it is more critical to achieve a good trade-off between privacy and model accuracy in this case. Our privacy budget allocation techniques optimize this trade-off by dynamically adjusting $\sigma$ during the training to improve model accuracy while retaining the same privacy guarantee.

\subsection{Evaluating Dynamic Privacy Budget Allocation}
\label{sec:budgetallocation}
In this section we evaluate the effectiveness of dynamic privacy budget allocation compared to uniform privacy budget allocation adopted by Abadi et al.~\cite{Abadi:2016:DLD}. Since the TensorFlow implementation uses random reshuffling to generate batches, privacy accounting in Section \ref{ssec:privacyRF} should be used to avoid the underestimation of privacy loss. We therefore use $\rho$ as the metric to represent the privacy budget and loss. Because the techniques for adjusting noise scales are independent of the batching method, the benefit of dynamic privacy budget allocation on model accuracy demonstrated under random reshuffling also applies to random sampling.

\subsubsection{Datasets and Models}
Our experiments use three datasets and different default neural networks for each dataset.

\noindent\textbf{MNIST}. This is a dataset of handwritten digits consists of 60,000 training examples and 10,000 testing examples~\cite{MNIST} formatted as 28X28 size gray-level images.
In our experiment, the neural network model for MNIST follows the settings in previous work~\cite{Abadi:2016:DLD} for comparison: a 60-dimensional PCA projection layer followed by a simple feed-forward neural network comprising a single hidden layer of 1000 ReLU units. The output layer is softmax of 10 classes corresponding to the 10 digits. The loss function computes cross-entropy loss. A batch size 600 is used. The non-private training of this model can achieve 0.98 accuracy with 100 epochs.

%\item{Iris}. The Iris dataset contains 150 rows of data that comprise 50 examples from each of three Iris species: Iris setosa, Iris virginica, and Iris versicolor. Each row contains sepal length and width, petal length and width, and flower species for each flower example. We use 120 rows as the training examples and 30 rows as the testing examples.

%We construct a neural network classifier and train it on the Iris data set to predict flower species based on sepal/petal geometry. We use the neural network model from TensorFlow tutorial~\cite{irisnetwork} that has three hidden layers, containing 10, 20, and 10 hidden ReLU units with bias terms, respectively. 

\noindent\textbf{Cancer Dataset}.
This dataset~\cite{cancerdataset} consists of 699 patient examples. Each example has 11 attributes including an id number, a class label that corresponds to the type of breast cancer (benign or malignant), and the 9 features describing breast fine-needle aspirates. After excluding 16 examples with missing values, we use 560 examples for training and 123 examples for testing. A neural network classifier with 3 hidden layers, containing 10, 20, and 10 ReLU units, is trained to predict whether a breast tumor is malignant or benign. Each iteration takes the whole training data set as a batch and thus each iteration is one epoch. The non-private training of this model achieves testing and training accuracy 0.96 with 800 epochs.

% \noindent\textbf{Census Income Dataset}.
% This dataset~\cite{censusincome} has 48842 records with 14 attributes such as age, workclass, education, occupation and working hours.
% The prediction task is to determine whether a person makes over 50K a year based on the census attributes. The data was preprocessed by excluding the examples with missing values and discretization of some continuous features. After preprocessing, each example has 39 features and 32561 examples are used for training and 16282 examples for testing.
% We construct a neural network classifier that has three hidden layers, each containing 256, 128, and 32 hidden ReLU units with bias terms respectively. The batch size is 200. The non-private training with learning rate 1e-4 achieves testing accuracy 0.79 and training accuracy 0.79 with 16 epochs.

\noindent\textbf{CIFAR-10}. The CIFA-10 dataset consists of 32$\times$32 color images with three channels (RGB) in 10 classes including ships, planes, dogs and cats. Each class has 6000 images. There are 40,000 examples for training, 10,000 for testing and 10,000 for validation.
For experiments on CIFAR, we use a pre-trained VGG16 neural network model~\cite{DBLP:journals/corr/SimonyanZ14a}. Following the previous work~\cite{Abadi:2016:DLD}, we assume the non-private convolutions layers that are trained over a public dataset (ImageNet~\cite{imagenet} for VGG16) and only retrain a hidden layer with 1000 units and a softmax layer with differential privacy. We use 200 training epochs and batch size of 200. The corresponding non-private training achieves 0.64 training accuracy and 0.58 testing accuracy.

\newcolumntype{b}{>{\hsize=1.3\hsize}X}
\newcolumntype{s}{>{\hsize=.6\hsize}X}
\newcolumntype{m}{>{\hsize=1\hsize}X}
\begin{table*}\small
\caption{Budget Allocation Schedules under Fixed Budget $\rho_{total}=0.78125$, initial noise scale $\sigma_0=10$ for dynamic schedules}
\centering
\begin{tabularx}{\linewidth}{|b|s|s|s|s|m|b|}
\hline
&Uniform\cite{Abadi:2016:DLD}~($\sigma_c$ =8) & Time ($k$=0.05) & Step($k$=0.6, $period$=10) & Exp ($k$=0.01) & Poly($k$=3,$\sigma_{end}$=2, $period$=100) & Validation($k$=0.7,$m$=5, $\delta$=0.01,$period$=10)\\
\hline
epochs & 100 & 38 & 31 & 71& 44 & 64 \\
\hline
training accuracy & 0.918& 0.934&0.928  &0.934 &0.930 & 0.930\\
\hline
testing accuracy & 0.919 & 0.931& 0.929 &0.929 &0.932 & 0.930\\ 
\hline
non-private SGD & 0.978/0.970 & 0.959/0.957& 0.955/0.954 &0.971/0.965 &0.963/0.959 & 0.97/0.964 \\
\hline
uniform/same \#epochs &  & 0.922/0.925 & 0.921/0.925 & 0.922/0.925 & 0.925/0.929 & 0.924/0.926 \\
\hline
\end{tabularx}
\label{tab:gt1}\vspace{-12pt}
\end{table*}

\subsubsection{Results on MNIST}
\label{ssec:mnist}
In differentially private model training, we keep the batch size at 600, clip the gradient norm of every layer at 4, and use fixed noise scale $\sigma_{pca}=16$ for differentially private PCA. Note that, since the PCA part has constant privacy cost $1/(2\sigma_{pca}^2)$ in terms of $\rho$-zCDP, we exclude it from the total privacy budget $\rho_{total}$ in our experiment, i.e., $\rho_{total}$ is only for the DP-SGD in Algorithm \ref{alg:dpsgd}.    
A constant learning rate 0.05 is used by default.
We evaluate the model accuracy during training under different privacy budget allocation schedules. For the validation-based schedule, we divide the training dataset into 55,000 examples for training and 5000 examples for validation, and perform validation every epoch. 

The results in Table \ref{tab:gt1} demonstrate the benefit of dynamic privacy budget allocations and the effect of earlier training termination on the model accuracy. 
For comparison, we consider the uniform privacy budget allocation in~\cite{Abadi:2016:DLD} with a constant noise scale $\sigma_c$=8 for every epoch as our baseline. We choose a fixed total privacy budget $\rho_{total}$=0.78125. This results in 100 training epochs in the baseline case.
We test all dynamic schedules with the hyperparameters given in the table and present their testing and training accuracy in numbers rounded to two decimals. The training is terminated when the privacy budget runs out, and the hyerparameters are chosen from a set of candidates to demonstrate varied training times in epochs which are reported in the table.
We also ran a non-private version of SGD with using the same training time as these schedules to see the impact of DP-SGD on accuracy.
We can see from Table \ref{tab:gt1} that the baseline with constant $\sigma_c$=8 achieves 0.918 training accuracy and 0.919 testing accuracy. By comparison, all non-uniform privacy budget allocation schedules improve the testing/training accuracy by 1\%$\sim$1.6\% while running fewer epochs. Because DP-SGD is a randomized procedure and the numbers in the table vary among trials, we repeat all the experiments 10 times and report in Figure \ref{fig:accuracy} the mean accuracy along with the min-max bar for every schedule. These results show that dynamic schedules consistently achieve higher accuracy than the baseline. Therefore, given a fixed privacy budget, the dynamic budget allocation can achieve better accuracy than using the uniform budget allocation.

The accuracy improvement shown with dynamic schedules in the above example comes from two sources: less training time and non-uniformity of budget allocation (i.e., decaying of the noise scale).
With uniform allocation under a fixed privacy budget, reducing the training time increases the privacy budget allocated to every epoch and thus decreases the noise scale used by the Gaussian mechanism. Therefore, when the model training benefits more from the reduced perturbation rather than longer training time, using less training time can improve the model accuracy. As verification, we apply uniform budget allocation with the noise scale $\sqrt{T/(2\rho_{total})}$ to achieve the same training time $T$ as the corresponding dynamic schedule. The training accuracy and testing accuracy are presented respectively in the final row of Table \ref{tab:gt1}. All cases outperform the baseline case with $\sigma_c=8$ with less than 100 training epochs, indicating the benefit of trading the training time for more privacy budget per epoch. however, it is worth noting that in certain cases, increasing the noise scale to prolong the training time  may help improve the accuracy, exemplified by the result of the validation-based schedule on the Cancer dataset.
Overall, when compared with the uniform allocation under the same training time, dynamic schedules demonstrate higher accuracy, therefore illustrating the benefit of non-uniformity and dynamic budget allocation.

\begin{figure}[!t]
    \centering
    \includegraphics[width=0.3\textwidth,height=0.2\textwidth]{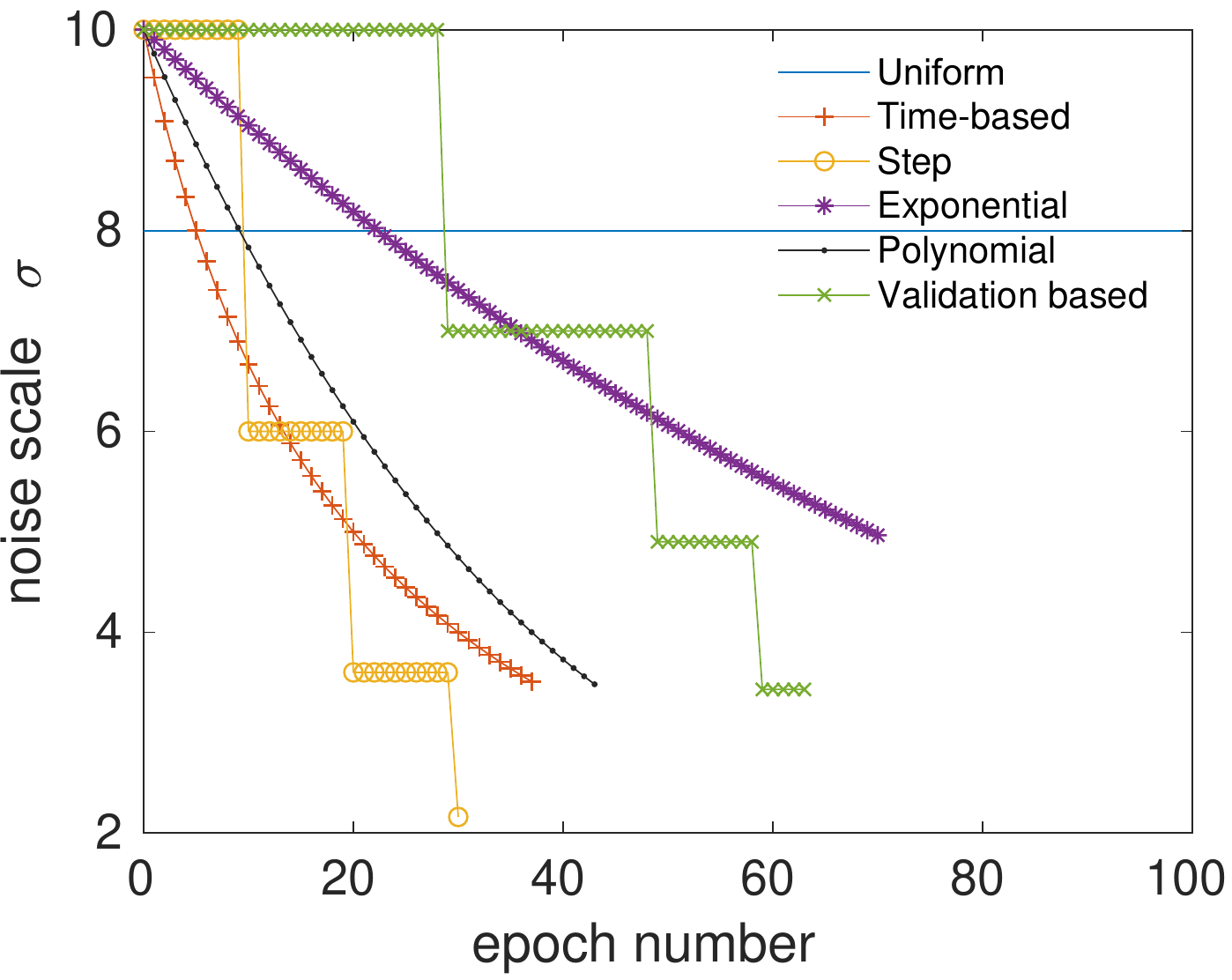}
    \caption{The change of noise scale $\sigma$ during training}
    \label{fig:schedules}\vspace{-10pt}
\end{figure}  
Figure \ref{fig:schedules} shows how the noise scale $\sigma$ changes with the epochs under different schedules in Table \ref{tab:gt1}. The curves terminate at the end of the training due to the depleted privacy budget. For the validation-based decay, the duration of the noise scale keeping unchanged decreases over the training time. The noise scale keeps 10 for 29 epochs, 7 for 20 epochs and 4.9 for 10 epochs. It is because that, as the model converges, the increment rate of the validation accuracy declines and it is more often to find that the accuracy increment does not exceed the given threshold.

We additionally manipulate different hyperparamters individually while keeping the rest constant to demonstrate their effects on training/testing accuracy and training time.. By default all accuracy numbers are the average of five trials.

\noindent\textbf{The effect of decay functions.}~
In our previous experiments, we evaluate four types of decay functions for the pre-defined schedules. Here we compare their effects on the model accuracy with constant training time. Given the total privacy budget and initial noise scale, we can use the composition theorem of $\rho$-zCDP to search for proper values of the parameter $k$ in these functions for the schedule to achieve a target training time. Table \ref{tab:gtk} in Appendix \ref{ssec:gtk} provides the values of $k$ for different decay functions to achieve training times of 60, 70, 80, 90, and 100 epochs respectively. These values are derived via search in step size $1e-4$, with the same initial noise scale and other parameters as stated in Table~\ref{tab:gt1}.

\begin{figure}
\centering
\includegraphics[width=0.4\textwidth,height=0.18\textwidth]{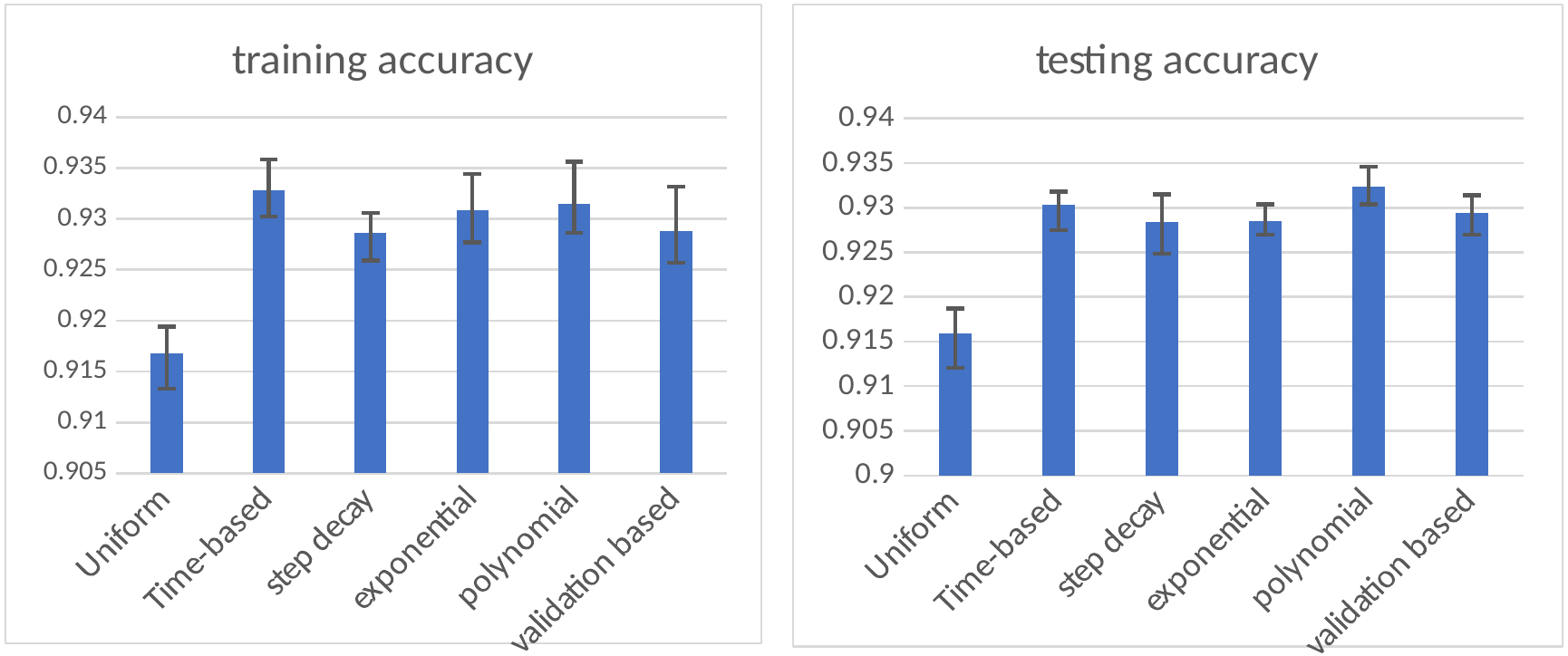}
\caption{The accuracy comparison of different schedules}
\label{fig:accuracy}
\includegraphics[width=0.5\textwidth,height=0.2\textwidth]{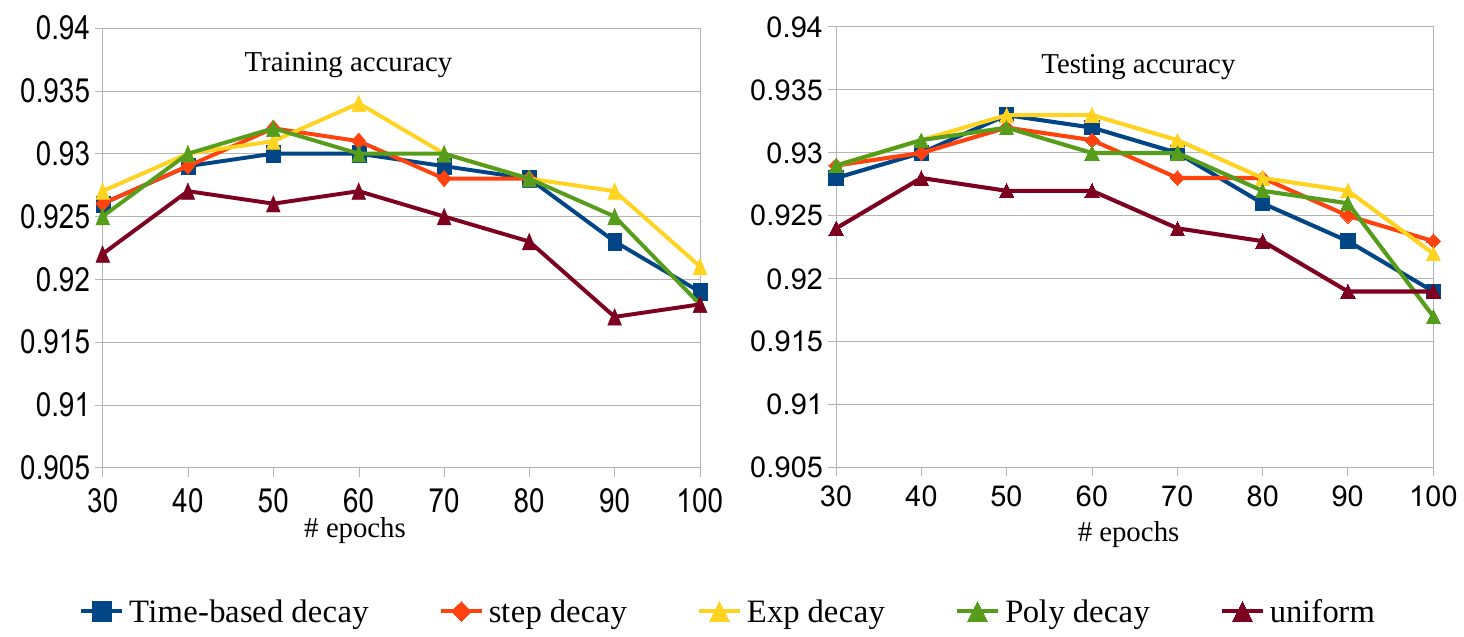}
\caption{The accuracy under fixed training time}
\label{fig:accuracyfixedtime}\vspace{-14pt}
\end{figure}

Figure \ref{fig:accuracyfixedtime} shows the training and testing accuracy of pre-defined schedules under different training times along with the accuracy achieved by uniform budget allocation~\cite{Abadi:2016:DLD} using the same privacy budget. We observe that all training instances using pre-defined schedules achieve higher accuracy compared to the uniform budget allocation given a fixed training time. However, there is no clear winner among the different decay functions. Their accuracy increases from 30 to 50 or 60 epochs and then decreases as the training time increases from 50 or 60 to 100 epochs. At 100 epochs, all pre-defined schedules have an accuracy closer to that of the uniform budget allocation schedule running 100 epochs. Due to the similar behaviors of the different decay functions, we choose to simply set the decay function to the exponential decay in subsequent experiments unless otherwise stated.

\begin{figure*}[!htb]
\minipage{0.25\textwidth}
  \includegraphics[width=\linewidth, height=0.7\linewidth]{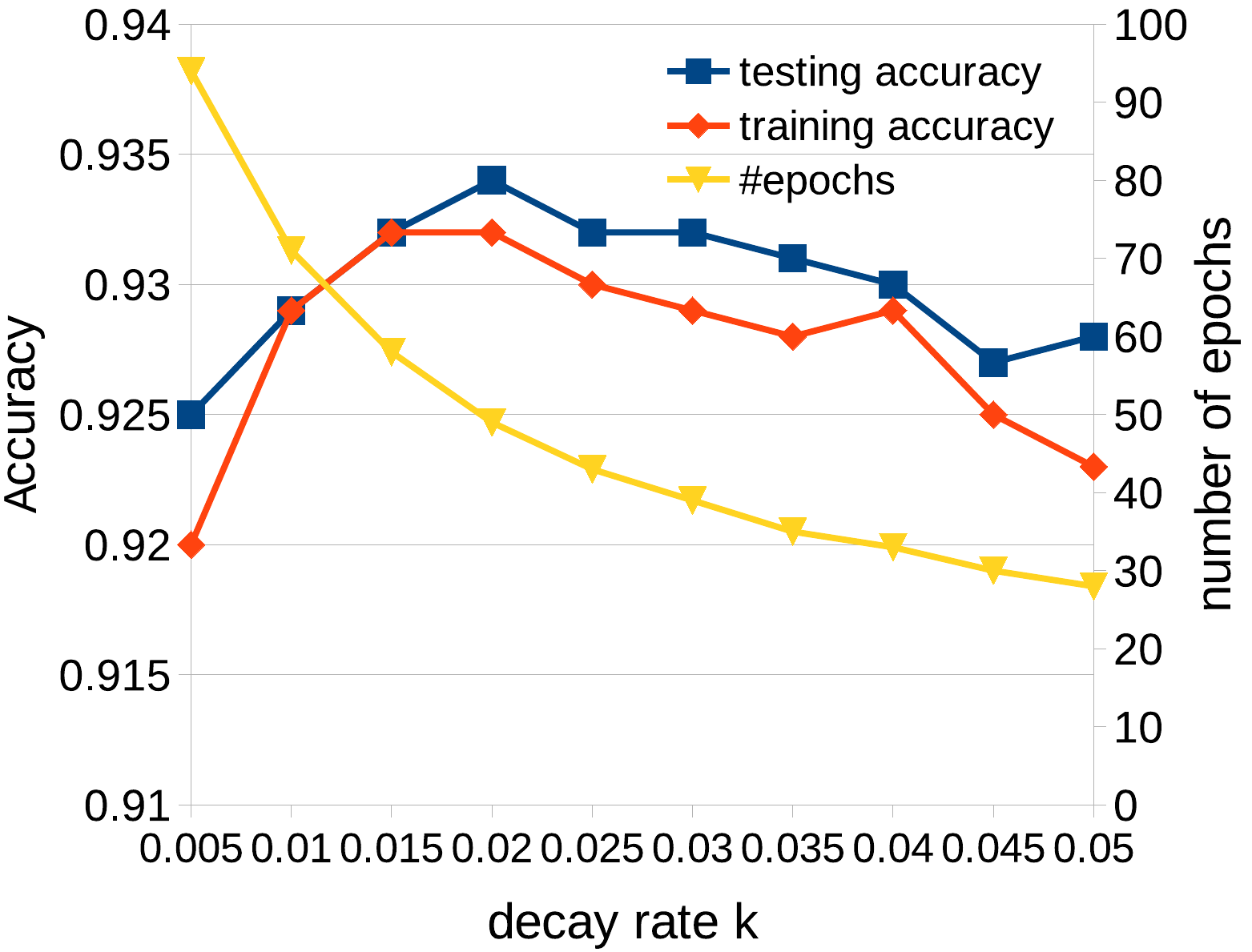}
  \caption{exponential decay}\label{fig:decayrate}
\endminipage\hfill
\minipage{0.25\textwidth}
  \includegraphics[width=\linewidth, height=0.7\linewidth]{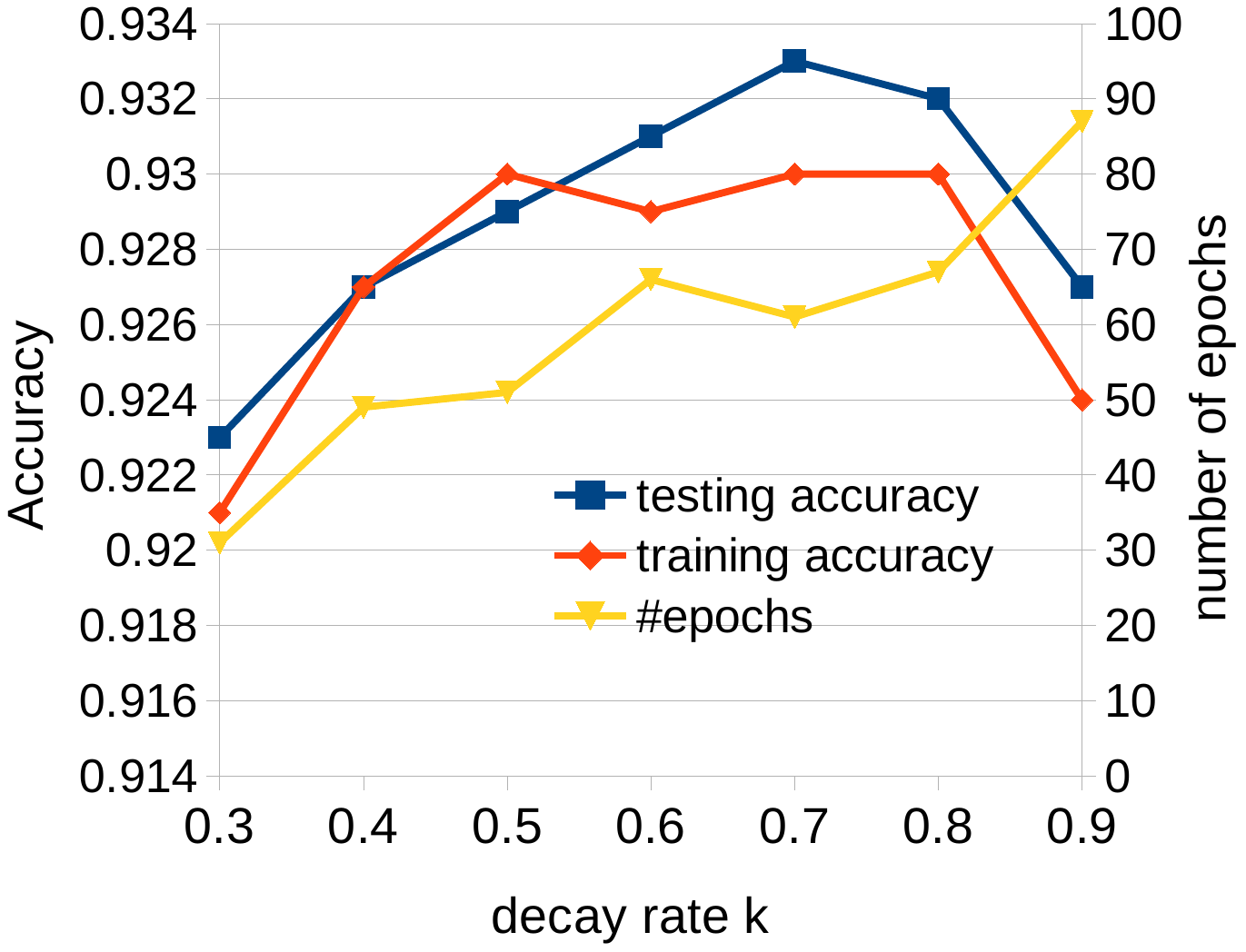}
  \caption{validation-based}\label{fig:vlidationrate}
\endminipage\hfill
\minipage{0.25\textwidth}%
  \includegraphics[width=\linewidth, height=0.7\linewidth]{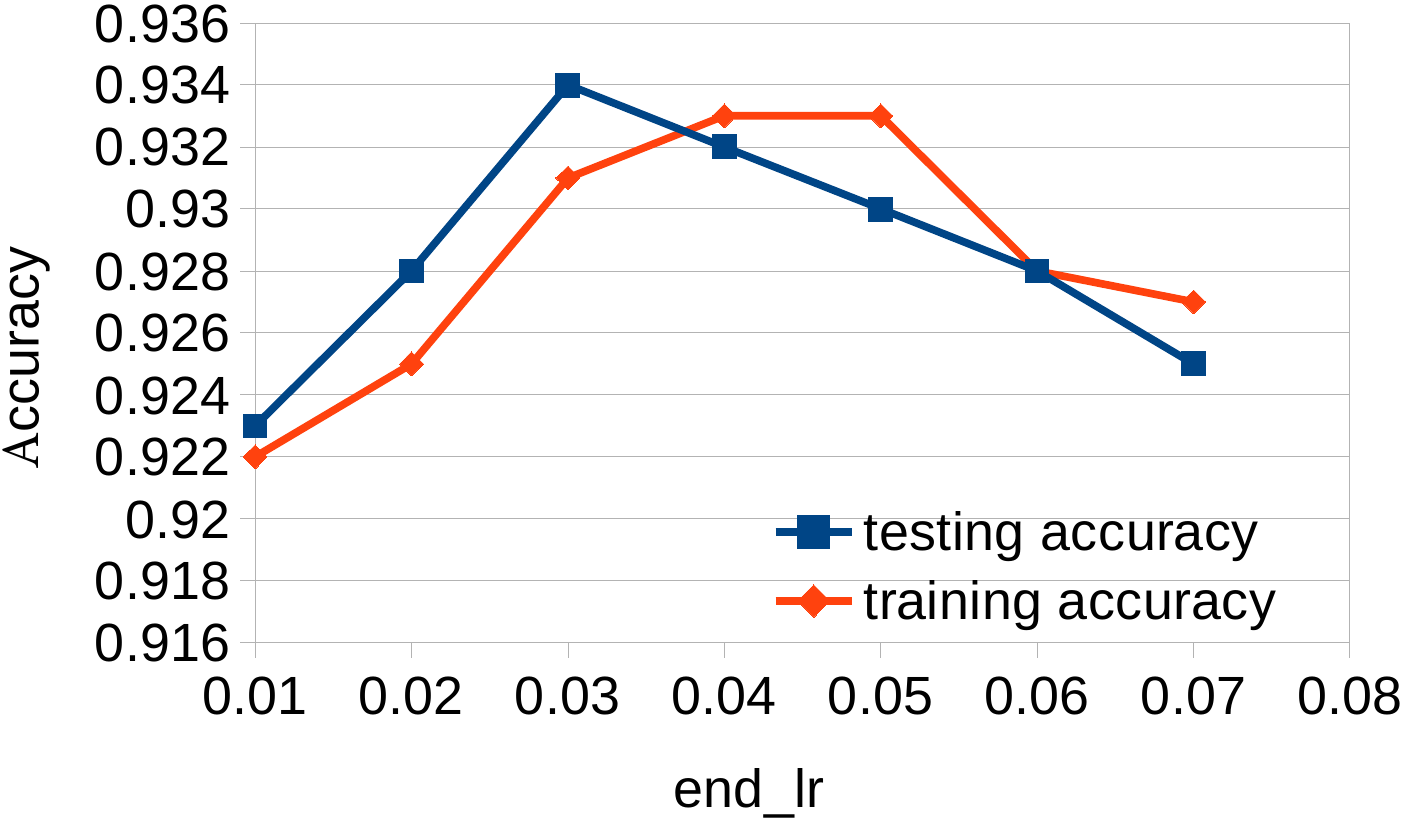}
  \caption{learning rate}\label{fig:learningrate}
\endminipage\hfill
\minipage{0.25\textwidth}%
  \includegraphics[width=\linewidth, height=0.7\linewidth]{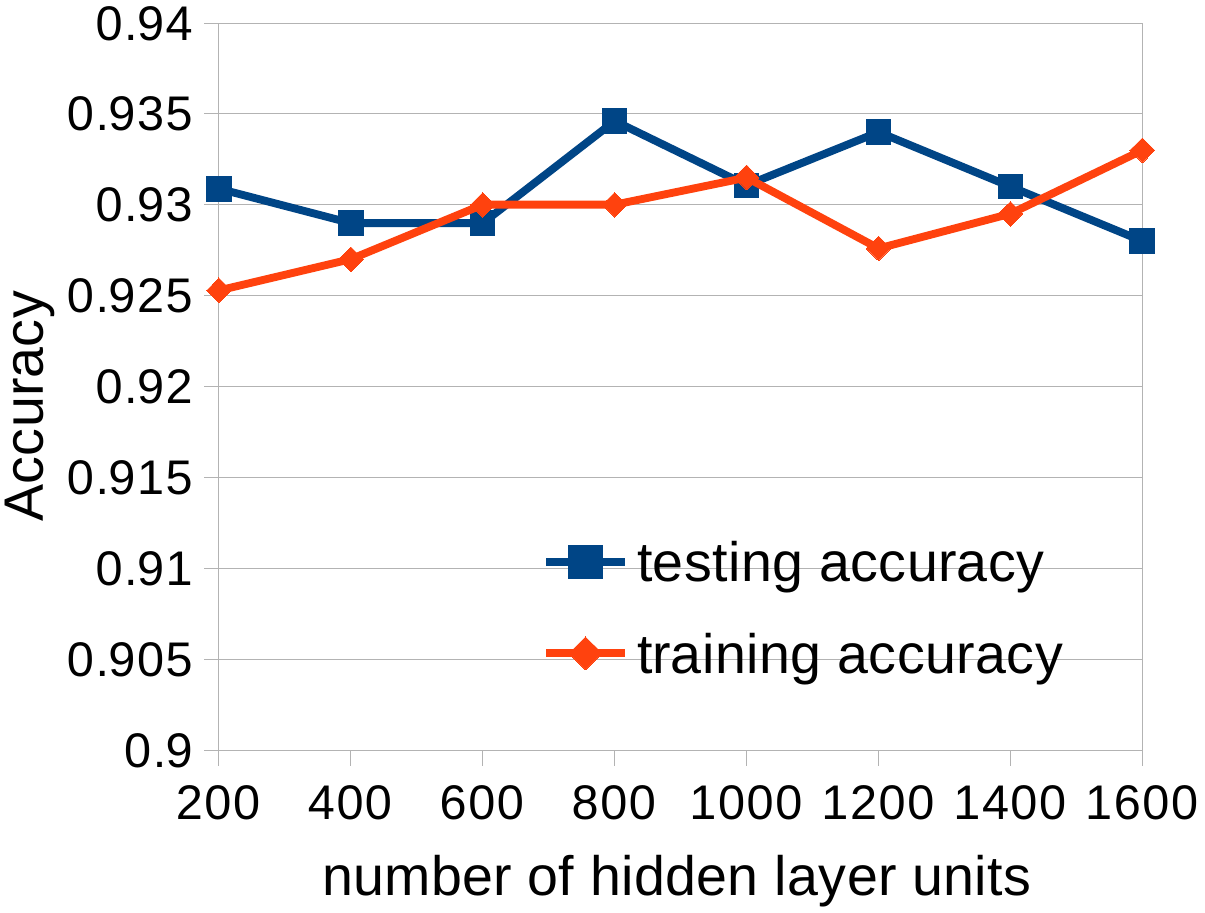}
  \caption{hidden units}\label{fig:hiddenunits}
\endminipage
\vspace{-10pt}
\end{figure*}

\noindent\textbf{Decay rate.}~
The decay rate $k$ decides how fast the noise scale decays. Keeping other parameters as the same as those reported in Table \ref{tab:gt1}, we vary $k$ from 0.005 to 0.5 for exponential decay and from 0.3 to 0.9 for validation-based decay. Figure \ref{fig:decayrate} and \ref{fig:vlidationrate} show the accuracy and training time under different values of $k$. We make three observations. First, in both cases, there exists an optimal decay rate to achieve maximum accuracy. For exponential decay, the best accuracy occurs at $k$=0.2; for validation-based decay, it occurs at $k$=0.7. Second,
for exponential decay, with the increase of $k$, the noise scale decreases at a higher rate. The privacy budget is therefore spent faster and the training time strictly decreases. For the validation-based decay, it reduces the noise scale to a $k$ fraction of the original, so the decay rate is actually 1-$k$ and the training time should increase with $k$. Figure \ref{fig:vlidationrate} shows that the training time overall increases with $k$ but with some random fluctuations.
This is because that the validation-based decay adjusts the noise scale according to the validation accuracy which may change during the training in a non-deterministic way.

Third, the results at the ends of the x-axis in both figures indicate two interesting facts. At one end, although the lowest decay rate leads to the longest training time, it also produces the worst accuracy because the noise scale decays more slowly in this case and therefore more epochs will suffer relatively higher noise scales. This degrades the efficiency of the learning process and lowers the accuracy. At the other end of the axis, the highest decay rate causes the training to stop much earlier, resulting in an insufficient training time which also degrades the accuracy.

\noindent\textbf{Learning rate.}~
Next we fix the decay rate $k$=$0.0138$ for exponential decay. Given a privacy budget 0.78125, the training lasts 60 epochs. With
an initial learning rate 0.1, we linearly decrease the learning rate to $end_{lr}$ over 10 epochs and then fix it at $end_{lr}$ thereafter. We vary $end_{lr}$ from 0.01 to 0.07. Figure \ref{fig:learningrate} shows that the accuracy decreases significantly when the learning rate is too small or too large. 

\noindent\textbf{Number of hidden units/layers.}
Next, we vary the number of hidden units in the model from 200 to 1600. The results are shown in Figure \ref{fig:hiddenunits}. Although more hidden units increase the sensitivity of the gradient, leading to more noise added at each iteration, we observe that increasing the number of hidden units does not decease the model accuracy under the exponential decay schedule. This is consistent with the observation in~\cite{Abadi:2016:DLD} using uniform budget allocation.
This shows that the effectiveness of dynamic budget allocation schedules scales to neural networks of different sizes. 
We also vary the number of hidden layers from 1 to 3, each with 1000 hidden units. The accuracy results are given in Appendix \ref{Appendix:layer} and are consistent with \cite{Abadi:2016:DLD} under the uniform allocation wherein the authors claim that, for MNIST, one hidden layer combined with PCA works better than networks with more layers. 

\begin{figure}[!t]
\centering
\begin{subfigure}{0.22\textwidth}
\includegraphics[width=1\textwidth, height=0.7\textwidth]{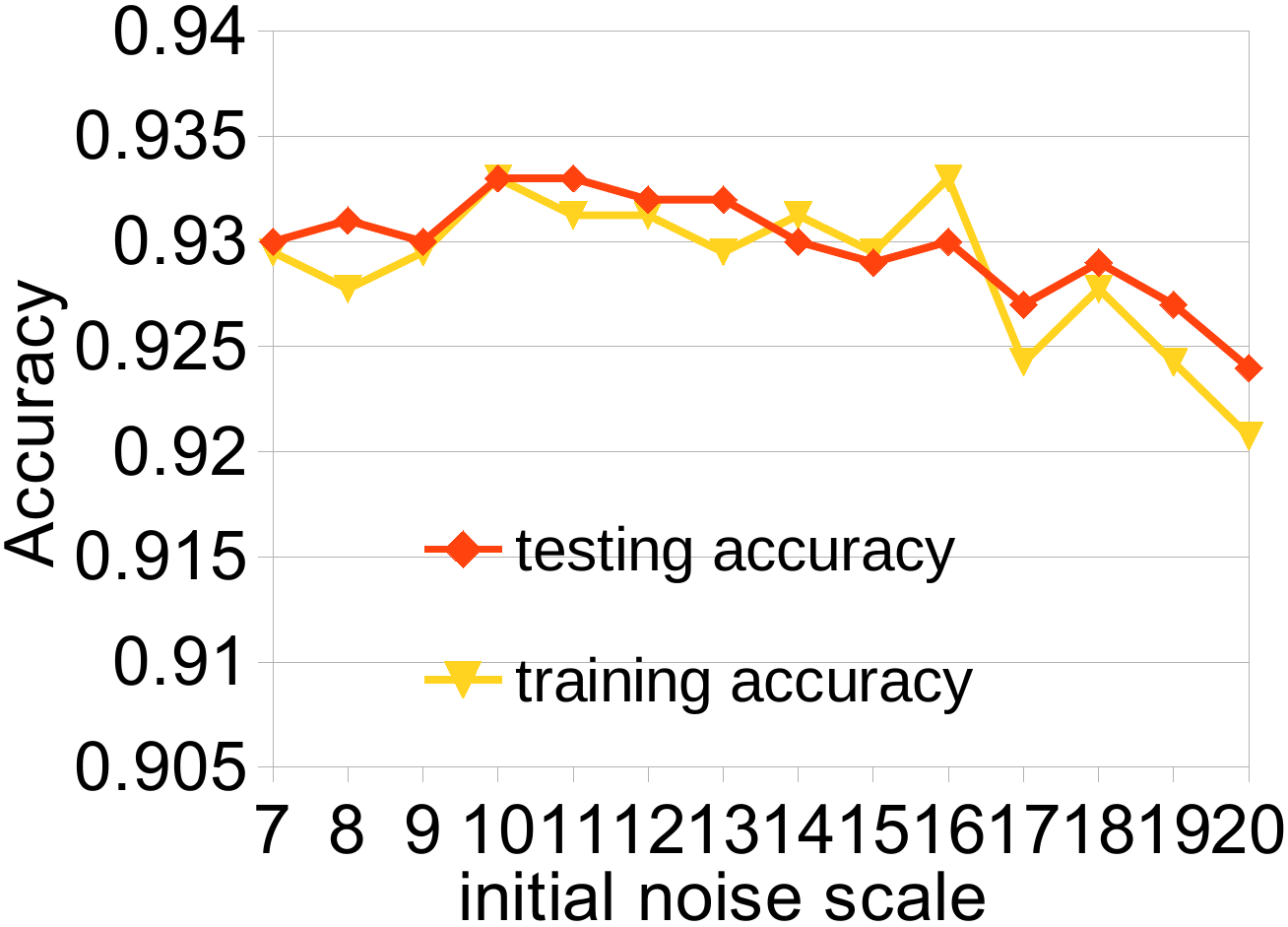}
    \caption{ fixed training time}
    \label{fig:init1}
\end{subfigure}
\begin{subfigure}{0.22\textwidth}
    \centering
    \includegraphics[width=1.1\textwidth,height=0.7\textwidth]{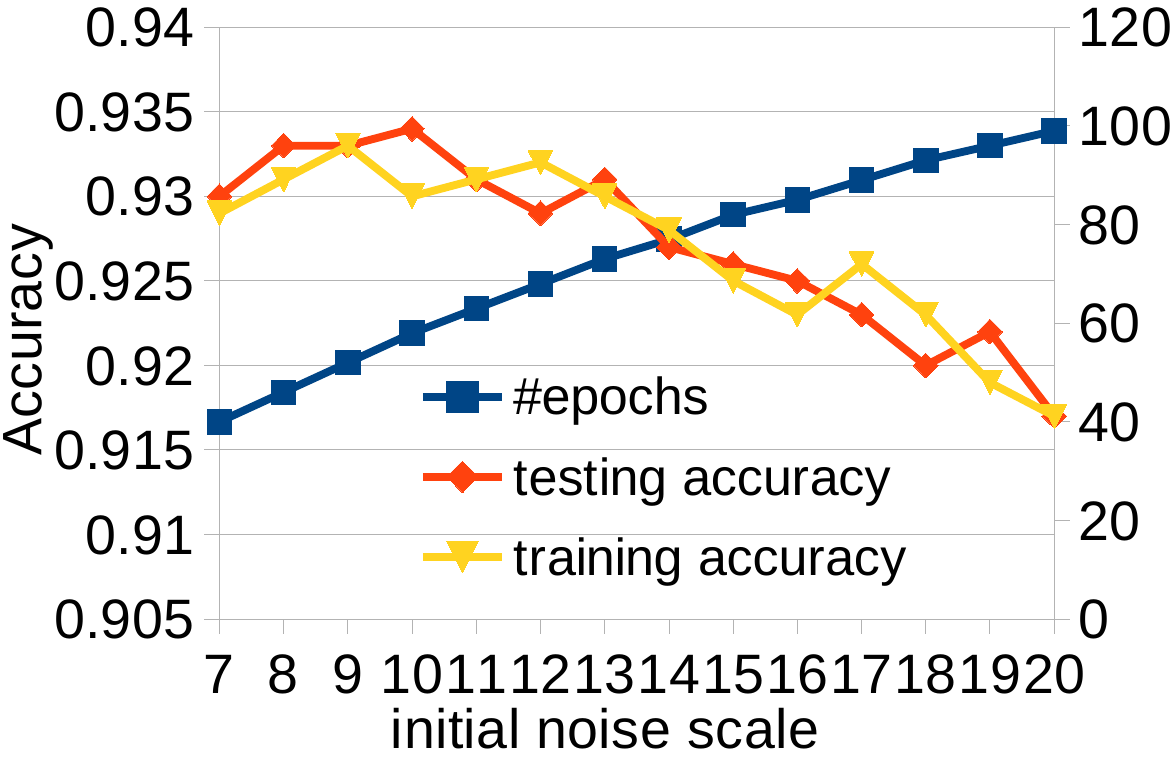}
    \caption{fixed decay rate}
    \label{fig:init2}
\end{subfigure}
\caption{Initial noise scale}
\minipage{0.25\textwidth}
  \includegraphics[width=0.9\linewidth,height=0.6\textwidth]{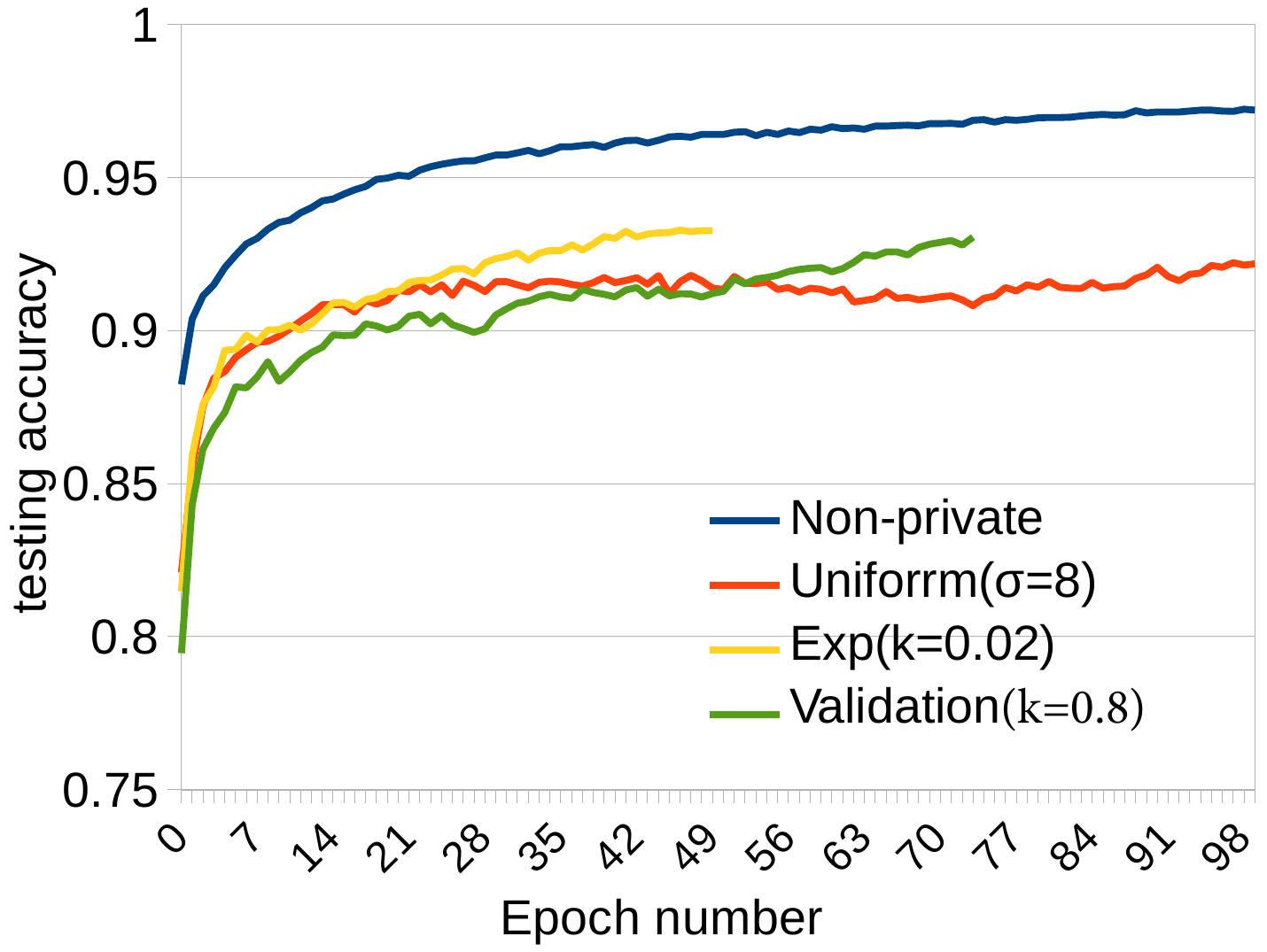}
  \caption{Accuracy in training}\label{fig:mnisttrend}
\endminipage
\minipage{0.25\textwidth}
  \includegraphics[width=0.9\linewidth,height=0.6\textwidth]{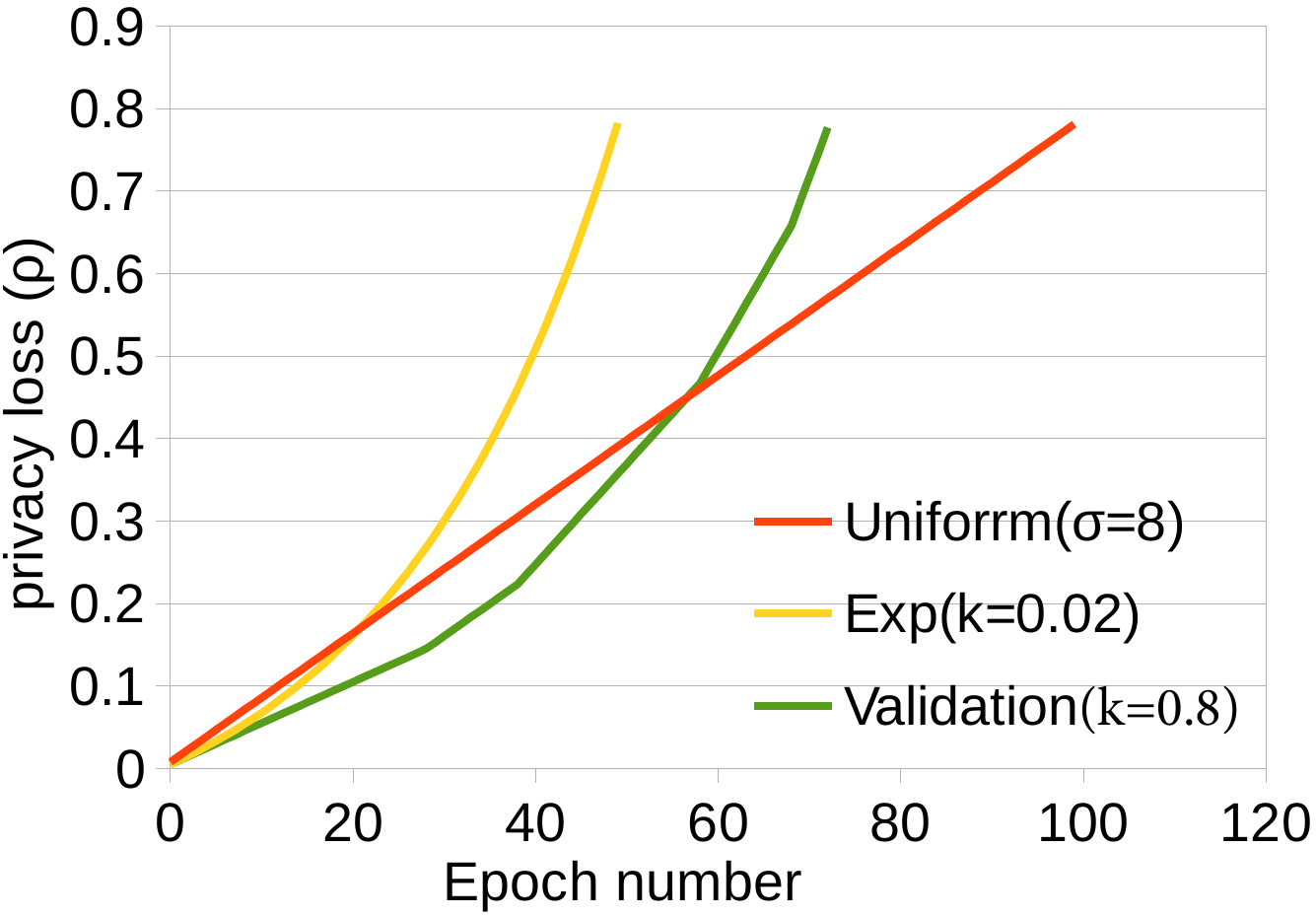}
  \caption{Privacy in training }\label{fig:privacytracking}
  \endminipage
  
\minipage{0.25\textwidth}
  \includegraphics[width=0.9\linewidth,height=0.7\linewidth]{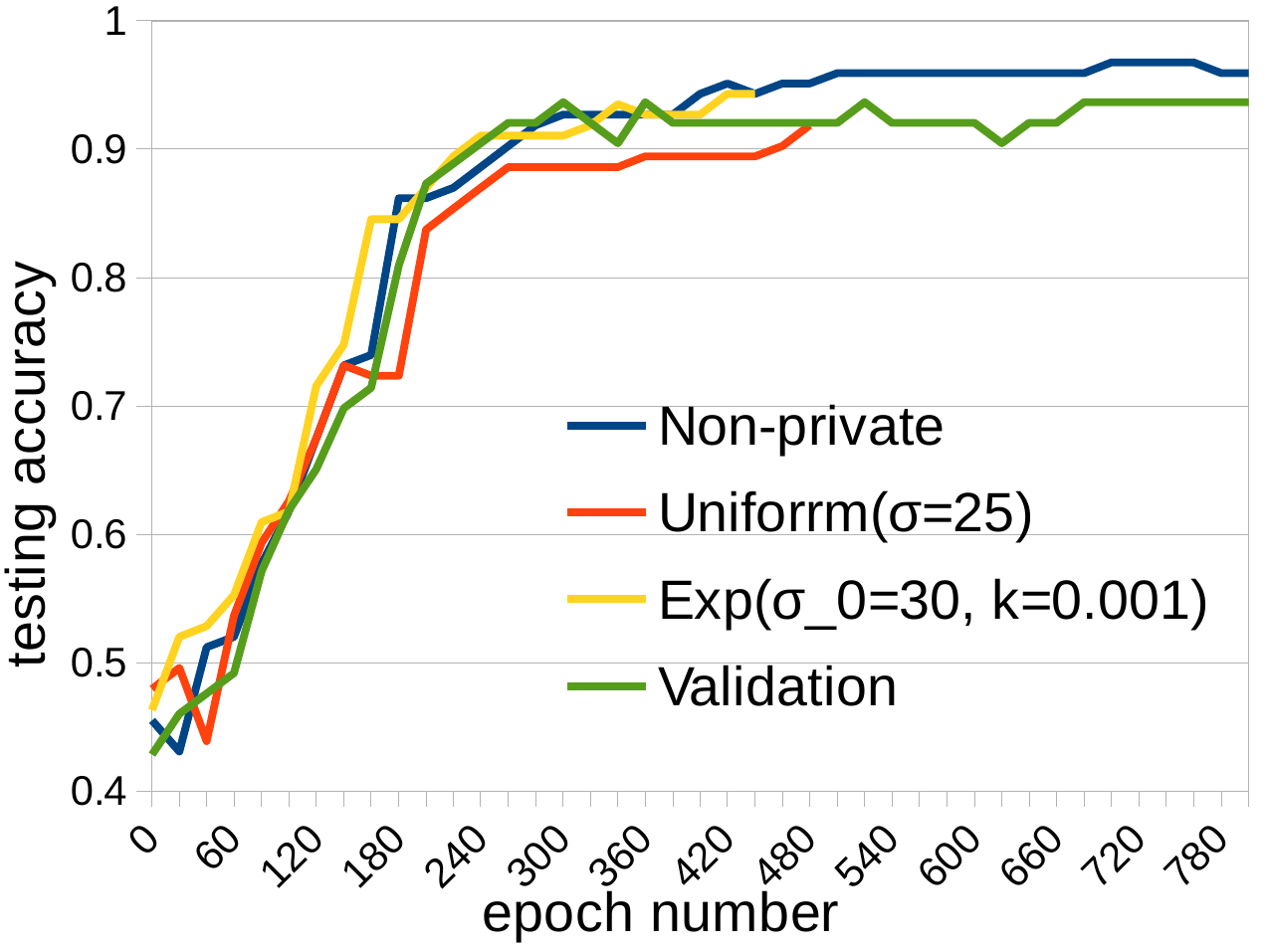}
  \caption{Accuracy (Cancer)}\label{fig:cancertrend}  \vspace{-13pt}
\endminipage~
\minipage{0.25\textwidth}
  \includegraphics[width=0.87\linewidth,height=0.7\linewidth]{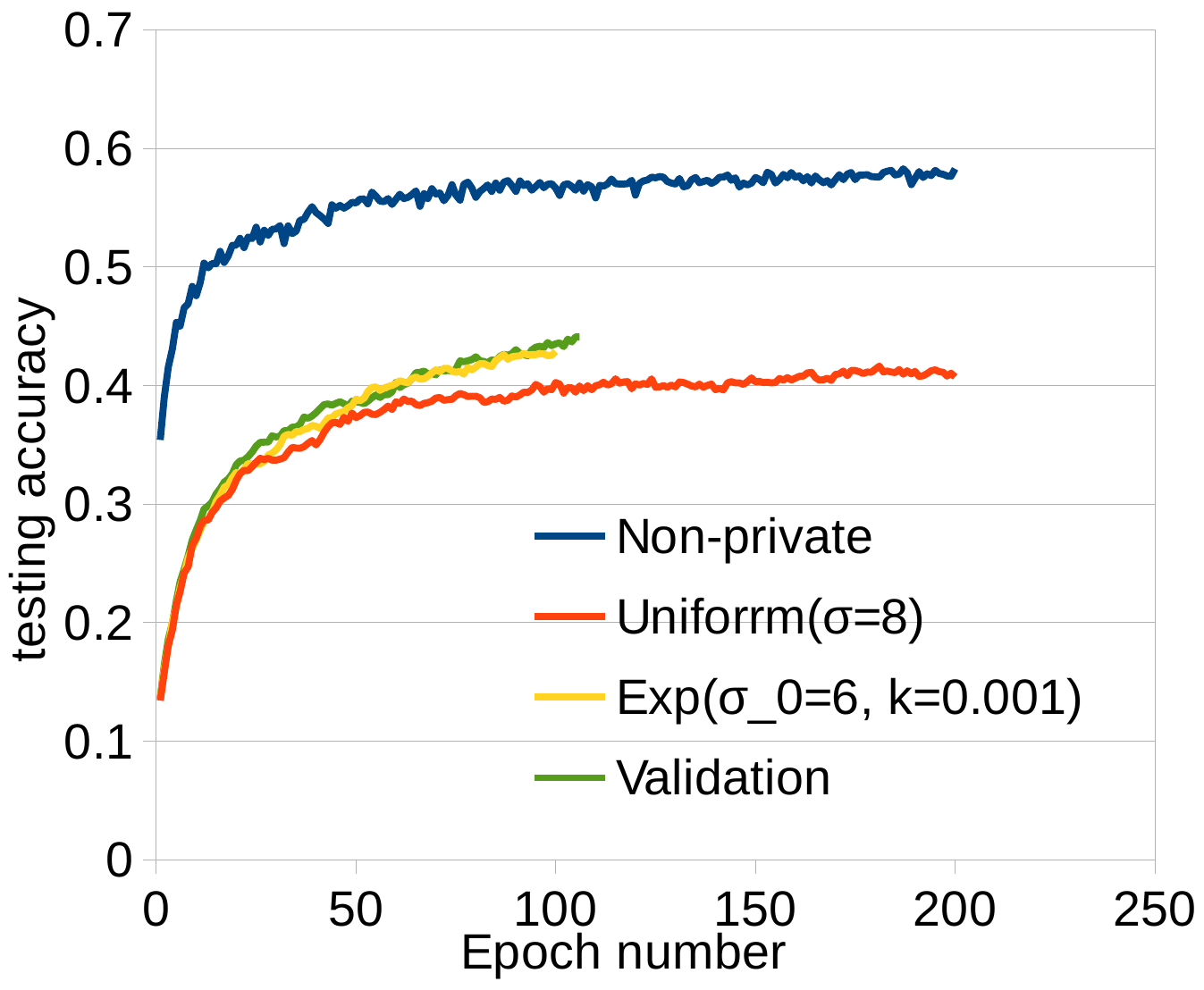}
  \caption{Accuracy (Cifar-10) }\label{fig:cifartrend}  \vspace{-13pt}
  \endminipage
\end{figure}

\noindent\textbf{Initial noise scale.}~
In the previous experiments, the initial noise scale $\sigma_0$=$10$ is set as the default. To examine the effect of $\sigma_0$, we vary its value from 7 to 20 and measure the model accuracy under the fixed privacy budget 0.78125 in two cases: 1) for each $\sigma_0$, we choose the exponential decay rate to achieve a fixed training time of 60 epochs; 2) the exponential decay rate is fixed to 0.015, leading to variation in training time. Figure \ref{fig:init1} and \ref{fig:init2} show that, overall, increasing the initial noise scale reduces accuracy. In comparing the two figures, we observe that when the training time is fixed, the choice of $\sigma_0$ has less impact on accuracy. This is because, when the training time is fixed, a larger $\sigma_0$ results in a higher decay rate of the noise scale which benefits accuracy. However, for fixed decay rate, although higher $\sigma_0$ leads to more training epochs, there is no accuracy improvement. This indicates that the model accuracy is more sensitive to the noise scale than the training time. 

\noindent\textbf{Accuracy and privacy in training.}~Figure \ref{fig:mnisttrend} and \ref{fig:privacytracking} illustrate the change of model accuracy and privacy loss during training time for schedules with the same parameters as in Table \ref{tab:gt1} except the parameters explicitly noted in the figures. Figure \ref{fig:privacytracking} shows that the uniform privacy budget allocation in~\cite{Abadi:2016:DLD} incurs linear growth of privacy loss in terms of $\rho$-zCDP while our dynamic budget allocation schedules have faster growth rate due to the reduction of noise scale with time. All instances stop when the given total privacy budget of 0.78125 is reached.
Combined with Figure \ref{fig:mnisttrend}, we can see that the exponential decay schedule consistently achieves better accuracy before the training ends compared to the uniform allocation, thanks to its faster noise scale reduction, while the validation-based schedule performs more conservatively and has a relatively longer training time.

An important implication of Figure \ref{fig:mnisttrend} is that the gap between uniform budget allocation and non-private SGD indicates the maximum potential for the accuracy improvement through dynamic budget allocation over the uniform allocation. The proposed dynamic budget allocation schedules provide users a way to improve the accuracy of DP-SGD to approach that of non-private SGD. It is not possible for dynamic budget allocation to completely close this gap because gradient perturbation inevitably hurts model accuracy. Therefore, we argue that the effectiveness of dynamic privacy budget allocation should be evaluated on how much it can reduce the gap between non-private SGD and DP-SGD with uniform allocation.
In our experiment, the accuracy difference between non-private SGD and DP-SGD of uniform case is 0.05 at the end of training. The exemplified schedules reduce this difference by 20\%$\sim$30\%. One of our ongoing research directions is to investigate the ways to effectively find the best hyperparameters to apply these schedules.

\subsubsection{Results on other datasets}
We repeat the experiments on the Cancer Dataset and CIFAR-10 datasets.
By applying exponential decay and validation-based decay to each learning task, we compare corresponding model accuracy with the uniform allocation method~\cite{Abadi:2016:DLD}. In this set of experiments, we first consider a uniform schedule that uses a constant noise scale to achieve a desired training time under the given privacy budget. Then we choose a value around this noise scale as the initial noise scale for decay schedules. A set of candidates for the decay rate is evaluated, and we use each candidate to train a model and compare achieved model accuracy. The parameters for the schedules we used are given in Table \ref{tab:gt30} in Appendix \ref{Appendix:s}.

Results for testing and training accuracy are show in Figures \ref{fig:cancertrend} and \ref{fig:cifartrend}. For the Cancer dataset, the exponential decay produces the model accuracy closer to the non-private SGD,  about 3\% higher accuracy than the uniform allocation case, and reduces the gap between non-private SGD and DP-SGD with uniform allocation by 70\%. The validation-based schedule produces about 1.8\% higher accuracy than the uniform case, with taking advantage of a longer training time as shown in Figure \ref{fig:cancertrend}. For CIFAR-10, the exponential decay achieves 2\% higher accuracy than the uniform case, and reduces the gap by about 12\%. The validation-based schedule improves model accuracy by 4\% over the uniform case, and reduces the gap by about 19\%.

% \begin{table}
% \caption{Results on datasets}
% \centering
% \small
% \begin{tabularx}{\linewidth}{|X|m|m|m|}
% \hline
% Dataset & Uniform & Exp Decay  & Validation\\
% \hline
% Cancer & 0.939/0.915 & 0.955/0.920 & 0.945/0.914      \\
% \hline
% CIFAR-10 & 0.702/0.534 & 0.718/0.54& 0.708/0.538 \\ 
% \hline
% \end{tabularx}
% \label{tab:gt3}
% \end{table}

\section{Discussion}
\label{sec:discussion}

We discuss a number of nuances/caveats as take-away remarks for deploying differentially private deep learning in practice for model publishing.

\noindent\textbf{Understanding privacy parameter.} Although differential privacy (DP) as a theory has evolved through different forms,
today it is still not clear how a realistic privacy benefit can be realized as a function of the privacy parameters in the DP definitions
such as the $\epsilon$ and $\delta$ parameters in traditional DP and the $\rho$ in zCDP. These privacy parameters lack understandable interpretations to the end-users. For $\rho$-zCDP, results like Proposition \ref{prop:zcdpvsdp} would help if $\epsilon$ and $\delta$ had straightforward privacy-related interpretations. Advancement in interpretability and usability of DP parameters by end-users and domain-scientists can have profound impact on the practical deployment of differential privacy.

\noindent\textbf{Data Dependency.}
The characteristics of input data, for example, dependency among training instances or dependency in the presence of training instances can render a differentially private mechanism ineffective for protecting the privacy of individuals~\cite{Kifer:2011:NFL,Hitaj:2017:DMU,DBLP:journals/corr/abs-1710-06963}. 
The baseline definition of differential privacy is focused on the privacy of a single instance and therefore when multiple instances of the same user are present, a DP mechanism needs to be extended to group-level differential privacy to provide sufficient protection. One direction of our future work is to investigate and explore the ways of extending our DP-SGD techniques to provide a group-level privacy guarantee.

\noindent\textbf{Resilience to Privacy Risks and Attacks.}
Differentially private deep learning aims to compute model parameters in a differentially private manner to limit the privacy risk associated with output model parameters. There are a number of known attacks in deep learning such as model inversion attacks and membership inference attacks. Model inversion attacks exploit the prediction output along with model access to infer an input instance. Membership inference attacks exploit the black box access to the prediction API to infer the membership of individual training instances. However, there is no formal study on whether or not a differentially private deep learning model is resilient to such attacks and what types of privacy risks known in practice can be protected with high certainty by a differentially private DNN model. In fact, DP only absolves the differentially-private release as a (quantifiably) strong cause of an inference. The work~\cite{DBLP:journals/corr/GhoshK16} provides an upper bound on the inferential privacy guarantee for differentially private mechanisms. DP, however, does not prevent the inference. This is another grand challenge in differential privacy and data privacy in general.

\section{Related Work}
\label{sec:related}

\noindent\textbf{Privacy threats in machine learning}
Existing works~\cite{Pharmacogenetics,Fredrikson:2015:MIA,Shokri:MI:2017,Song:2017:MLM} have shown that machine learning models and their usage may leak information about individuals in the training dataset and input data. Fredrikson, et al.~\cite{Pharmacogenetics} proposed a model inversion attack, which uses the output/prediction produced by a model to infer the unknown features of the input data and apply this attack against decision trees and neural networks in a pharmacogenetics scenario~\cite{Fredrikson:2015:MIA}. Reza et al.~\cite{Shokri:MI:2017} developed a membership inference attack that aims to determine if an individual record was used as part of the training dataset for the model using only the black-box access to the target model. Song et al.~\cite{Song:2017:MLM} proposed training phase attacks which perform minor modifications to training algorithms to make them output models which encode a significant amount of information about the training dataset while achieving high quality metrics like accuracy and generalizability. In addition, model extraction attacks proposed in~\cite{stealml} aim to duplicate the functionality of the model with black-box access. Such attacks can be leveraged to infer information about the model's training dataset.

\noindent\textbf{Privacy-preserving deep learning}
To enable deep learning over the data from multiple parties while preserving the privacy of each party's training dataset, Reza et al ~\cite{Shokri:2015:PDL} proposed a distributed deep learning framework that lets the participants train their model independently on their own dataset and only selectively share a subsets of their models' parameters during training. Abadi et al.~\cite{Abadi:2016:DLD} proposed a differentially private SGD algorithm for deep learning to offer provable privacy guarantees on the output model. DP~\cite{dip} as a defacto standard for privacy has been applied to various machine learning algorithms, such as logistic regression~\cite{Chaudhuri:2008:PLR,Zhang:2012:FMR}, support vector machines~\cite{Rubinstein2009LearningIA} and risk minimization~\cite{Chaudhuri:2011:DPE,BassilyST14}, aiming to limit the privacy risk associated with the output model parameters on the training dataset. Our work in this paper is primarily related to~\cite{Abadi:2016:DLD}. We improve their approach in a number of ways. For example, instead of using traditional $(\epsilon, \delta)$-differential privacy, we apply concentrated differential privacy~\cite{Dwork2016ConcentratedPrivacy,BunS16} to provide tight cumulative privacy loss estimation over a large number of computations. Furthermore, we characterize the effect of data batching methods on the composition of differential privacy and propose a dynamic privacy budget allocation framework for improving the model accuracy.

\section{Conclusion}
\label{sec:conclude}
We have presented our approach to differentially private deep learning for model publishing with three original contributions. First, since the training of neural networks involves a large number of iterations, we apply CDP for privacy accounting to achieve tight estimation on privacy loss. Second, we distinguish two different data batching methods and propose privacy accounting methods for privacy loss estimation under each respectively. Third, we have implemented several dynamic privacy budget allocation techniques for improving model accuracy over existing uniform budget allocation schemes. Our experiments on multiple datasets demonstrate the effectiveness of dynamic privacy budget allocation.

\section*{Acknowledgment}
The authors would like to thank our anonymous reviewers for their valuable comments and suggestions.
This research was partially sponsored by NSF under grants SaTC 156409, CISE’s SAVI/RCN (1402266, 1550379), CNS (1421561), CRISP (1541074), SaTC (1564097) programs, an REU supplement (1545173), an RCN BD Fellowship, provided by the Research Coordination Network (RCN) on Big Data and Smart Cities, an IBM Faculty Award, and gifts, grants, or contracts from Fujitsu, HP, Intel, and Georgia Tech Foundation through the John P. Imlay, Jr. Chair endowment. Any opinions, findings, and conclusions or recommendations expressed in this material are those of the author(s) and do not necessarily reflect the views of the funding agencies and companies mentioned above.

\bibliographystyle{abbrv}
\bibliography{dp}
\clearpage

\newpage
\appendix
\setlength{\abovedisplayskip}{0pt}
\setlength{\belowdisplayskip}{0pt}
\renewcommand{\theorem}{Theorem}
\renewcommand{\property}{Property}
\renewcommand{\corollary}{Corollary}

\subsection{Proof of Theorem \ref{recompose}}
\begin{theorem}{ \ref{recompose}.}
Suppose that a mechanism $\mathcal{A}$ consists of a sequence of $k$ adaptive mechanism $\mathcal{A}_1,\ldots, \mathcal{A}_k$ where each $\mathcal{A}_i:~ \prod_{j=1}^{i-1}\mathcal{R}_j \times \mathcal{D} \rightarrow \mathcal{R}_i$ and $\mathcal{A}_i$ satisfies $\rho_i$-zCDP. Let  $\mathbb{D}_1, \mathbb{D}_2, \ldots, \mathbb{D}_k$ be the result of a randomized partition of the input domain $\mathbb{D}$. 
The mechanism $\mathcal{A}(D) = (\mathcal{A}_1(D\cap \mathbb{D}_1), \ldots,\mathcal{A}_k(D\cap \mathbb{D}_k))$ satisfies\begin{equation}
\begin{cases}
\rho \text{-zCDP}, & \text{if } \rho_i=\rho, \forall i \\
\max\limits_i \rho_i \text{-zCDP},  & \text{if } \rho_i \neq \rho_j \text{ for some } i, j
\end{cases}
\end{equation}
\end{theorem}
\begin{IEEEproof}
Suppose two neighboring datasets $D$ and $D'$. Without loss of generality, assume that $D$ contains one more element $d_e$ than $D'$. Let $D_i= D \cap \mathbb{D}_i$ and $D'_i= D'\cap \mathbb{D}_i$. Accordingly, there exists $j$ such that $D_j$ contains one more element than $D'_j$, and for any $i \neq j$, $D_i=D'_i$.
Consider any sequence of outcomes $o = (o_1,\ldots,o_k)$ of $\mathcal{A}_1(D_1), \ldots, \mathcal{A}_k(D_k)$.

Because only $D_j$ is different from $D'_j$,
for any $i\neq j$, we have
$$\Pr[\mathcal{A}_i(D_i)=o_i|\mathcal{A}_{i-1}(D_{i-1})=o_{i-1}, \ldots, \mathcal{A}_{1}(D_{1})=o_{1}]$$  equal to
$$\Pr[\mathcal{A}_i(D'_i)=o_i|\mathcal{A}_{i-1}(D'_{i-1})=o_{i-1}, \ldots, \mathcal{A}_{1}(D'_{1})=o_{1}]$$

Then, we have
\begin{align}
&L^{(o)}_j \overset{\Delta}{=} \log  \bigg(\frac{\Pr(\mathcal{A}(D)=o}{\Pr(\mathcal{A}(D')=o)} \bigg) \notag \\
&= \log\bigg(\frac{\prod_{i \in [n]} \Pr[\mathcal{A}_i^{D_i}=o_i|\mathcal{A}_{i-1}^{D_{i-1}}=o_{i-1},.., \mathcal{A}_{1}^{D_{1}}=o_{1}]}
{\prod_{i \in [n]} \Pr[\mathcal{A}_i^{D'_i}=o_i|\mathcal{A}_{i-1}^{D'_{i-1}}=o_{i-1},.., \mathcal{A}_{1}^{D'_{1}}=o_{1}]} \bigg) \notag \\
&= \log\bigg(\frac{ \Pr[\mathcal{A}_j^{D_j}=o_j|\mathcal{A}_{j-1}^{D_{j-1}}=o_{j-1},.., \mathcal{A}_{1}^{D_{1}}=o_{1}]}
{\Pr[\mathcal{A}_j^{D'_j}=o_j|\mathcal{A}_{j-1}^{D'_{j-1}}=o_{j-1},.., \mathcal{A}_{1}^{D'_{1}}=o_{1}]} \bigg) \notag \\
& \overset{\Delta}{=} c_j(o_j; o_1,\ldots, o_{j-1}) \notag
\end{align}
where $\mathcal{A}_i^{D_i}$ denotes $\mathcal{A}_i {(D_i)}$ for short.

Once the prefix ($o_1,\ldots, o_{j-1}$) is fixed, 
$$C_j \overset{\Delta}{=} c_j(o_j; o_1,\ldots, o_{j-1}) = \log \bigg( \frac{\Pr(\mathcal{A}_j(D_j) = o_j)}{\Pr(\mathcal{A}_j(D'_j) = o_j)} \bigg)$$

By the $\rho_j$-zCDP property of $\mathcal{A}_j$, $\mathbb{E}\left[e^{(\alpha-1) C_j} \right] \le e^{(\alpha-1)\alpha \rho_j}$, thus
\begin{align*}
\mathbb{E}\left[e^{(\alpha-1) L^{(o)}_j} \right] &= \mathbb{E}\left[\exp \bigg((\alpha-1) C_j\bigg) \right] \\
&\le e^{(\alpha-1)\alpha \rho_j}
\end{align*}

Because of randomized partition of the input domain $\mathbb{D}$,
the extra element $d_e$ of $D$ is randomly mapped to $k$ partitions. Therefore, $j$ is uniformly distributed over $\{1, \ldots, k\}$, and thus the privacy loss $L^{(o)}$ under random data partition is the mixture of independent random variables $L^{(o)}_1, \cdots, L^{(o)}_k$, 

$$
f(L^{(o)}) = \frac{1}{k}f(L^{(o)}_1) + \ldots + \frac{1}{k}f(L^{(o)}_k)
$$
where $f(X)$ is the probability distribution function of $X$.

We have
\begin{align*}
\mathbb{E}\left[e^{(\alpha-1) L^{(o)} } \right]&  =\frac{1}{k} \sum_{j=1}^k  \mathbb{E}\left[ \exp \bigg((\alpha-1) L^{(o)}_j\bigg) \right]\\
\end{align*}

Because $L^{(o)}_j$ satisfies zCDP, by (\ref{eq:zcdpeq}) then we have
\begin{align*}
& \le \frac{1}{k}\sum_{j=1}^k \exp((\alpha-1)\alpha\rho_j) 
\end{align*}

If $\rho_j=\rho$ $\forall j$,  we have $\mathbb{E}\left[e^{(\alpha-1) L^{(o)} } \right] \le  \exp((\alpha-1)\alpha\rho)$, and thus
the mechanism $\mathcal{A}(D)$ satisfies $\rho$-zCDP.

If not all $\rho_j$ are the same, we replace each $\rho_j$ with $\max\limits_j \rho_j$, we have $\mathbb{E}\left[e^{(\alpha-1) L^{(o)} } \right] \le  \exp((\alpha-1)\alpha \max\limits_j \rho_j)$, and the mechanism $\mathcal{A}(D)$ satisfies $\max_i \rho_i $-zCDP.
\end{IEEEproof}

\subsection{Proof of Theorem \ref{recompose3}}
\begin{theorem}{ \ref{recompose3}. }
\label{recompose3_A}
Let $\widehat{\rho} =P(q,\sigma)$ and $u_\alpha = U_\alpha(q, \sigma)$.
If the mechanism $\mathcal{A'}$ has
\begin{equation}
D_\alpha (\mathcal{A'}(\mathbb{D})||\mathcal{A'}(\mathbb{D}')) \le \alpha \widehat{\rho}
\end{equation}
for $1<\alpha \le u_\alpha$, it satisfies
\begin{numcases}{}
\small
\hspace{-5pt}\Big(\widehat{\rho} + 2\sqrt{\widehat{\rho}\log(1/\delta)},\delta \Big)-DP, \rm{~if~} \delta \ge  1/\exp(\widehat{\rho}(u_\alpha-1)^2) &  \notag \\
\hspace{-5pt}\Big(\widehat{\rho}u_\alpha-\frac{\log \delta}{u_\alpha-1},\delta \Big)-DP, otherwise &  \notag
\end{numcases}
\end{theorem}
\begin{IEEEproof}

Let $Z=L^{(o)}_{(\mathcal{M}(D)||\mathcal{M}(D'))}$ be privacy loss random variable, then for $1<\alpha \le u_\alpha$,
\begin{equation}
\mathbb{E} \left [ e ^ {(\alpha-1)Z} \right] =e ^ {(\alpha-1)D_\alpha (M(\mathbb{D})||M(\mathbb{D}'))} \le e ^{(\alpha-1) \alpha \widehat{\rho} }
\end{equation}
By Markov' inequality,
\begin{align}
P(Z \ge \epsilon) = &P(e^{(\alpha-1)Z} > e^{(\alpha-1) \epsilon}) \le \frac{\mathbb{E} \left [ e ^ {(\alpha-1)Z} \right]}{e^{(\alpha-1) \epsilon}} \\
& \le \rm{exp}((\alpha-1) ( \widehat{\rho} \alpha - \epsilon))
\end{align}

The unconstrained minimum of function $g(\alpha)=(\alpha-1) (\widehat{\rho}\alpha - \epsilon)$ occurs at $\alpha^*=(\epsilon+\widehat{\rho})/(2\widehat{\rho}) $, and the minimum value is $-(\epsilon - \widehat{\rho})^2 /(4\widehat{\rho})$. 
If $\alpha^* \le u_\alpha$, this unconstrained minimum corresponds to the constrained minimum as well that is subject to $\alpha \le u_\alpha$. Let $\delta = \rm{exp}\big( -(\epsilon - \widehat{\rho})^2 /(4\widehat{\rho}) \big)$, and then we have $\epsilon=\widehat{\rho} + 2\sqrt{\widehat{\rho}log(1/\delta)}$, which has the same form as in Proposition \ref{prop:zcdpvsdp}. In this case,  $$\alpha^*=(2\widehat{\rho} + 2\sqrt{\widehat{\rho}log(1/\delta)})/(2\widehat{\rho})$$, and it requires

\begin{equation}
\label{eq:neq1}
( 2\widehat{\rho} + 2\sqrt{\widehat{\rho}log(1/\delta)})/(2\widehat{\rho} ) \le u_\alpha
\end{equation}
From (\ref{eq:neq1}) we have
\begin{equation}
\delta \ge 1/\exp(\widehat{\rho}(u_\alpha-1)^2)
\end{equation}

Otherwise, 
if 
\begin{equation}
\delta <  1/\exp(\widehat{\rho}(u_\alpha-1)^2)
\end{equation}
which means $$\alpha^* > u_\alpha$$
Then, because the function $g(\alpha)$ is monotonically decreasing in the interval $(0, \alpha^*]$, the constrained minimum is achieved at the boundary point $$\alpha^+ =u_\alpha$$ and accordingly we let $\delta=\rm{exp}\big( (\alpha^+-1)(\widehat{\rho} \alpha^+ -\epsilon) \big)$ and have $$\epsilon = \widehat{\rho}\alpha^+ - \frac{log\delta}{\alpha^+-1}$$.
\end{IEEEproof}

\subsection{$\rho$-zCDP mechanism $\mathcal{A}$ with random sampling still satisfies $\rho$-zCDP}
\label{ssec:samerho}
Suppose $\mathcal{A'}$ that runs $\rho$-zCDP mechanism $\mathcal{A}$ on a random subsample of dataset $\mathbb{D}$. Following the result (\ref{eq:stillrho}) in the paper, the proof is as follows:

Because $\Pr(\Lambda(\mathbb{D}')=T'|d_e \in T')=\Pr(\Lambda(\mathbb{D})=T)$ and $\mathcal{A}$ satisfies $\rho$-zCDP, R\'enyi divergence \begin{align}
D_\alpha(u_0||u_1)\le D_\alpha(\mathcal{A}(T)||\mathcal{A}(T'))\le \alpha \rho  \label{a:1}\\
D_\alpha(u_1||u_0)\le D_\alpha(\mathcal{A}(T')||\mathcal{A}(T))\le \alpha \rho
\end{align}
Using jointly quasi-convexity of R\'enyi divergence~\cite{Divergence}, we have
\begin{align}
&D_\alpha {(u_0||q u_1 +(1-q) u_0)} \le D_\alpha(u_0||u_1)  \le \alpha \rho  \label{ee1}\\
&D_\alpha {(q u_1 +(1-q) u_0~||~u_0)} \le D_\alpha(u_1||u_0)  \le \alpha \rho  \label{ee2} 
\end{align}
and thus $\mathcal{A'}$ still satisfies $\rho$-zCDP.

\begin{table*}[!t]
\caption{decay rate values for different training times}
\centering
\setlength\tabcolsep{1.5pt}
\begin{tabular}{|c|c|c|c|c|c|c|c|c|c|}
\hline
\#epochs & 30 &40 & 50 &60& 70& 80& 90 &100\\
\hline
Time & 0.076& 0.0441 & 0.0281 & 0.019 & 0.0132 & 0.0093 & 0.0067& 0.0048  \\
\hline
Step & 0.5459& 0.7008 & 0.7922 &0.851 & 0.891 &0.919 &0.94 &0.956 \\
\hline
Exp &0.0442 &0.0282 &0.0193 &0.0138 & 0.0101& 0.0075 &0.0056 &0.0041 \\ 
\hline
Poly &6.2077 & 3.5277 & 2.1948 & 1.4317 & 0.9549& 0.6382 &0.4167 &0.1626 \\
\hline
\end{tabular}
\label{tab:gtk}
\end{table*}

\subsection{The Impact of number of hidden layers for MNIST}
\label{Appendix:layer}
See Table \ref{tab:layers}.
\begin{table}[H]
\caption{Accuracy results under different number of layers}
\centering
\small
\begin{tabularx}{\linewidth}{|X|X|X|X|}
\hline
Layer numbers & 1 & 2 & 3\\
\hline
training accuracy & 0.9315 & 0.9228 &0.902\\
\hline
testing accuracy & 0.931 & 0.921 &  0.898 \\ 
\hline
\end{tabularx}
\label{tab:layers}
\end{table}

\subsection{Decay rate values for different training times}
\label{ssec:gtk}
Table \ref{tab:gtk} list the values of decay rate used for each decay function to achieve the corresponding training times.

\subsection{Schedule parameters for the Cancer and CIFAR-10 datasets}
\label{Appendix:s}
See Table \ref{tab:gt30}.

\begin{table}[H]
\caption{Schedule parameters}
\centering
\small
\begin{tabularx}{\linewidth}{|X|X|X|X|}
\hline
Dataset & Uniform & Exp Decay & Validation\\
\hline
Cancer & $\sigma$=25, \#epochs=500 & $\sigma_0$=30, $k$=0.001 & $\sigma_0$=35,$k$=0.99, $period$=50,$\delta$ =0.01, $m$=1\\
\hline
CIFAR-10 & $\sigma$=8, \#epochs=200 & $\sigma_0$=6, $k$=0.001 & $\sigma_0$=6,$k$=0.99, $period$=10,$\delta$ =0.05, $m$=5 \\ 
\hline
\end{tabularx}
\label{tab:gt30}
\end{table}

\subsection{Validation of emperical settings for bounding $\alpha$-R\'enyi divergence}
\label{Appendix:v}

 \begin{figure}[ht]
 \centering
 \begin{subfigure}{0.22\textwidth}
 \includegraphics[width=1\textwidth]{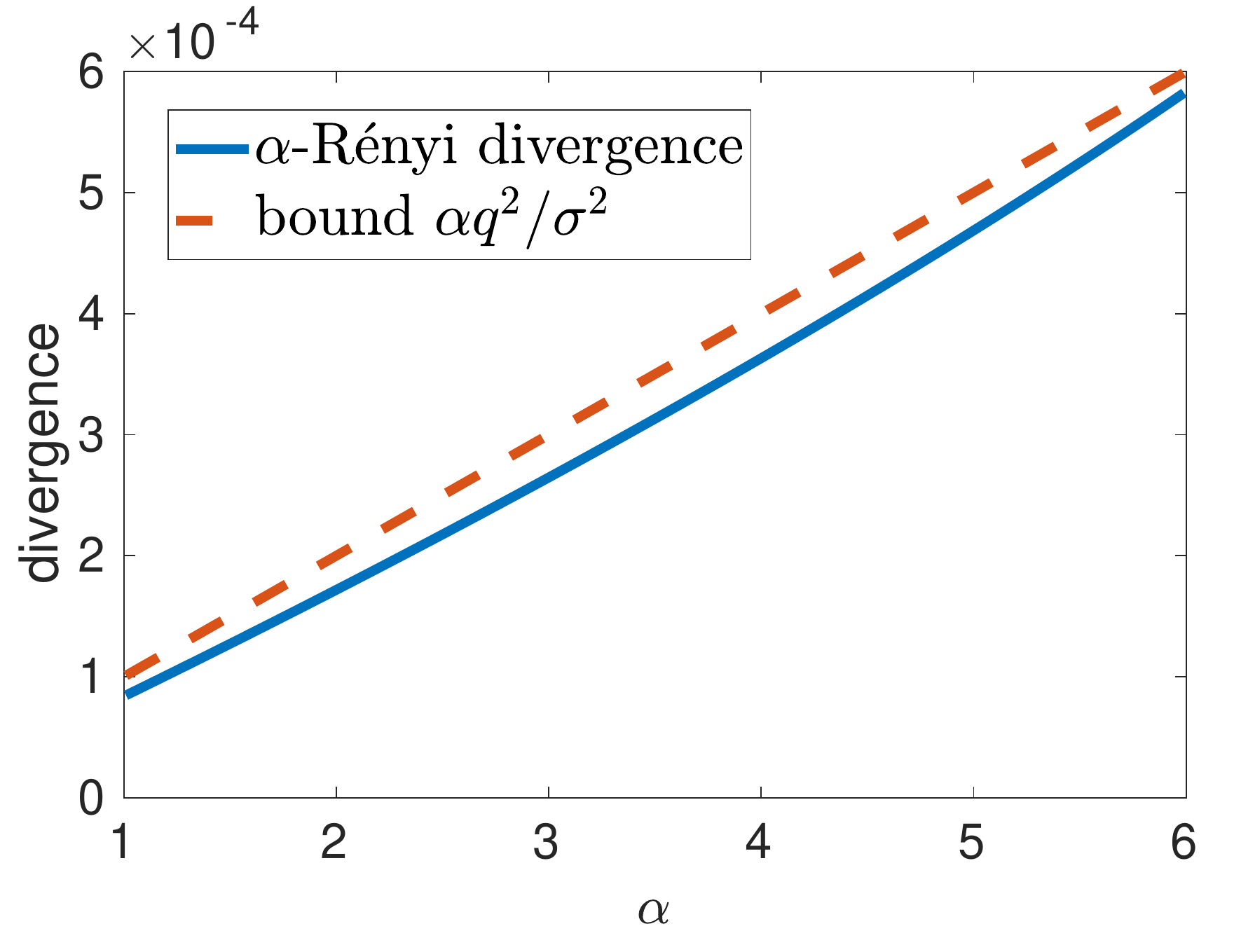}
     \caption{$\sigma=1,q=0.01$}
     \label{fig:valid2}
 \end{subfigure}
 \begin{subfigure}{0.22\textwidth}
     \centering
     \includegraphics[width=1\textwidth]{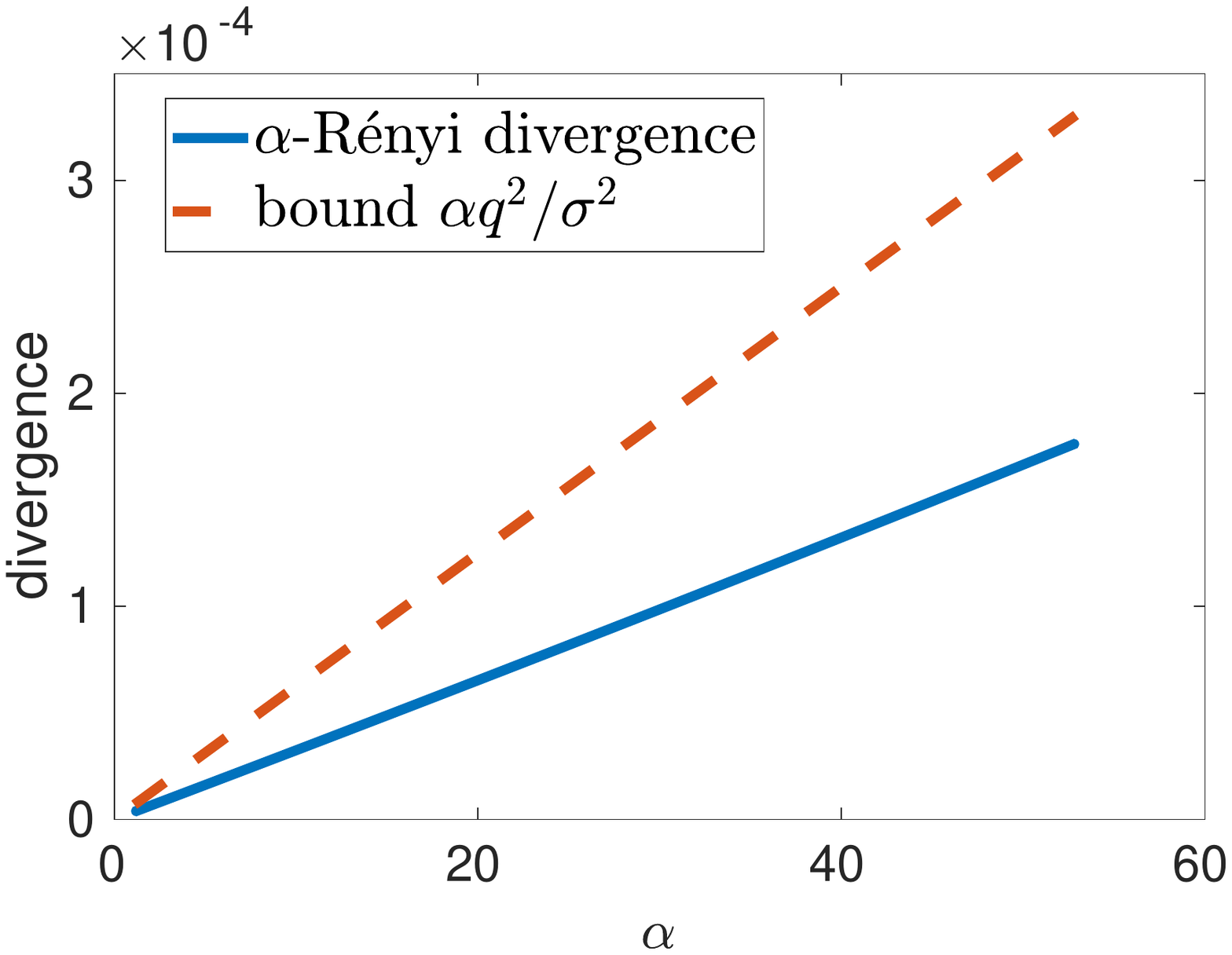}
     \caption{$\sigma=4,q=0.01$}
     \label{fig:valid1}
 \end{subfigure}
 \caption{Bound v.s. $\alpha$-R\'enyi divergence}
 \label{fig:cc}
 \end{figure}

\vspace{-10pt}

In this section we describe our numerical validation of the bound $P=q^2\alpha/\sigma^2$ of $\alpha$-R\'enyi divergence for the Gaussian mechanism with random sampling given the condition $q \le \frac{1}{16\sigma}$ and $\alpha$ $ \in (1, U_\alpha = \sigma^2\log \frac{1}{q\sigma}+1]$. We compute the $\alpha$-R\'enyi divergence with the code for moments accountant (named as Renyi differential privacy (RDP) analysis in Tensorflow)(https://github.com/tensorflow/privacy/blob/master/privacy/\\analysis/rdp\_accountant.py).

We test $\sigma$ and $q$ at a precision of 0.001, by varying $\sigma$ from 1 to 30 and $q$ from 0.001 to $\frac{1}{16\sigma}$ with a step size of 0.001, $\alpha$ $ \in (1, U_\alpha = \sigma^2\log \frac{1}{q\sigma}+1]$ with step size 0.01. 
In our validation, the bound $q^2\alpha/\sigma^2$ is close but always higher than the $\alpha$-R\'enyi divergence. We plot Figure \ref{fig:cc} for demonstration.
We also did randomly additional tests for cases such as a smaller step size and a wide $\sigma$ value up to 50. We found that the empirical bound $q^2\alpha/\sigma^2$ always holds under the given conditions. Therefore, we have the conjecture that $q^2\alpha/\sigma^2$ is a valid bound for the $\alpha$-R\'enyi divergence of sampled Gaussian mechanism and seek its formal proof in our future work. With using $P$ and $U_\alpha$ as stated above for Theorem \ref{recompose3}, we can have an effective estimation on the privacy loss in terms of $(\epsilon, \delta)$-DP in our tested parameter settings of $\sigma$ and $q$.

%\section*{Notes:}
%It was brought to our attention by some authors of CCS’16 paper~\cite{Abadi:2016:DLD} that the log moment bound $Tq^2\lambda^2/\sigma^2$ used in the proof of Theorem 1 in~\cite{Abadi:2016:DLD} should have an asymptotic term $O(q^3 \lambda^3 /\sigma^3)$ as in Lemma 3 in the appendix of~\cite{Abadi:2016:DLD}. Accordingly, in this paper we remove Lemma 1 of our conference paper that is derived "by applying the log moment bound $q^2\lambda^2/\sigma^2$ of the privacy" in the proof of Theorem 1 in~\cite{Abadi:2016:DLD}, and revise Theorem 3. We also correct an error in the statement of Theorem 2 of our conference paper~\cite{Yu2018DifferentiallyPM}.

\subsection{Code for testing Tensorflow batch APIs}
\label{ap:batch}
\small{
\begin{lstlisting}[float=*]
import numpy as np
import tensorflow as tf
data = np.array(range(10))
n = 1
limited_tensor = tf.train.limit_epochs(data, n)
batch_size = 1
batch = tf.train.shuffle_batch([limited_tensor], batch_size=batch_size, 
		enqueue_many=True, capacity=batch_size*100, min_after_dequeue=batch_size*10)
input_fn = tf.estimator.inputs.numpy_input_fn(data, batch_size=batch_size, 
		num_epochs=n, shuffle=True)
data_ops = input_fn()

#true random sampling with sampling ratio q = batch_size/input_size
data_size = data.shape[0]
q=batch_size/float(data_size)
print(q)
for step in range(data_size/batch_size):
    s = np.random.uniform(0,1,size=data_size)
    print(data[s<=q])


with tf.Session() as sess:
  sess.run(tf.initialize_local_variables())
  tf.train.start_queue_runners()
  print("test tf.train_shuffle_batch")
  try:
    while True:
      data_batch = sess.run(batch)
      print(data_batch)
      # process data
  except tf.errors.OutOfRangeError:
    pass

  print("test tf.estimator.inputs.numpy_inpunt_fn")
  try:
    while True:
      data_batch = sess.run(data_ops)
      print(data_batch)
      # process data
  except tf.errors.OutOfRangeError:
    pass
\end{lstlisting}
}

% %\subsection{proof}
% %\label{Appendix:z}
% %Suppose $\alpha_0=n$ and $\alpha_1=n+1$ where $n \in \mathcal{N}$. If $D_{\alpha_0}(P||Q) \le \alpha_0 q^2/\sigma^2$, $D_{\alpha_1}(P||Q) \le \alpha_1 q^2/\sigma^2$, let $\alpha_x = n+x$ where $0 < x < 1$,  we want to prove $D_{\alpha_x}(P||Q) \le \alpha_x q^2/\sigma^2$.
% %
% %We first show the log moment generating function is convex in its moment order.
% %Consider the log moment of variable $Z$, $\log E(\exp(((1-\theta)\lambda_0+\theta\lambda_1)Z))$ where $0 < \theta < 1$.
% %By the Holder's inequality $E(UV) \le (E|U|^p)^{1/p} (E|V|^q)^{1/q}$ for any $ 1 < p, q <\infty$ having $\frac{1}{p}+\frac{1}{q} =1$, 
% %we have
% %\begin{equation}
% %\begin{split}
% %&\log E(\exp(((1-\theta)\lambda_0+\theta\lambda_1)Z)) \\
% %& = \log E(\exp((1-\theta)\lambda_0)Z * \exp(\theta \lambda_1 Z) ) \\
% %&\text{let~} U=e^{((1-\theta)\lambda_0)Z}, V = e^{\theta \lambda_1 Z}, p = \frac{1}{1-\theta}, q = \frac{1}{\theta}, \text{then} \\
% %&\le (1-\theta) \log E(\exp(\lambda_0 Z)) + \theta \log E(\exp(\lambda_1 Z))
% %\end{split}
% %\end{equation}

\end{document}